\documentclass[iop]{emulateapj}

\usepackage{natbib}
\usepackage{hyperref}
\usepackage{graphicx}
\usepackage{multirow}

\newcommand{\msun}{\mbox{M}_{\odot}}
\newcommand{\mstar}{$\mbox{M}_*$}

\newcommand{\secpoint}{\mbox{$''\mskip-7.6mu.\,$}}

\shorttitle{The MOSDEF Survey: Electron Density and Ionization Parameter at $z\sim2.3$}
\shortauthors{Sanders et al.}

\begin{document}

\title{The MOSDEF Survey: Electron Density and Ionization Parameter at $\lowercase{z}\sim2.3$\altaffilmark{*}}
\altaffiltext{*}{Based on data obtained at the
W.M. Keck Observatory, which is operated as a scientific partnership among the California Institute of Technology, the
University of California, and NASA, and was made possible by the generous financial support of the W.M. Keck Foundation.
}

\author{Ryan L. Sanders\altaffilmark{1}} \altaffiltext{1}{Department of Physics \& Astronomy, University of California, Los Angeles, 430 Portola Plaza, Los Angeles, CA 90095, USA}

\author{Alice E. Shapley\altaffilmark{1}}

\author{Mariska Kriek\altaffilmark{2}} \altaffiltext{2}{Astronomy Department, University of California, Berkeley, CA 94720, USA}

\author{Naveen A. Reddy\altaffilmark{3,5}} \altaffiltext{3}{Department of Physics \& Astronomy, University of California, Riverside, 900 University Avenue, Riverside, CA 92521, USA}

\author{William R. Freeman\altaffilmark{3}}

\author{Alison L. Coil\altaffilmark{4}} \altaffiltext{4}{Center for Astrophysics and Space Sciences, University of California, San Diego, 9500 Gilman Dr., La Jolla, CA 92093-0424, USA}

\author{Brian Siana\altaffilmark{3}}

\author{Bahram Mobasher\altaffilmark{3}}

\author{Irene Shivaei\altaffilmark{3}}

\author{Sedona H. Price\altaffilmark{2}}

\author{Laura de Groot\altaffilmark{3}}

\altaffiltext{5}{Alfred P. Sloan Fellow}

\email{email: rlsand@astro.ucla.edu}

\begin{abstract}

Using observations from the MOSFIRE Deep Evolution Field (MOSDEF) survey, we investigate the physical conditions of star-forming
 regions in $z\sim2.3$ galaxies, specifically the electron density and ionization state.
  From measurements of the [O~\textsc{ii}]$\lambda\lambda$3726,3729 and [S~\textsc{ii}]$\lambda\lambda$6716,6731 doublets, we find a median
 electron density of $\sim250$~cm$^{-3}$ at $z\sim2.3$, an increase of an order of magnitude compared to measurements
 of galaxies at $z\sim0$.
  While $z\sim2.3$ galaxies are offset towards significantly higher O$_{32}$ values relative to local galaxies at fixed stellar mass,
 we find that the high-redshift sample follows a similar distribution to the low-metallicity tail of the local distribution
 in the O$_{32}$ vs. R$_{23}$ and O3N2 diagrams.
  Based on these results, we propose that $z\sim2.3$ star-forming galaxies have the same ionization parameter as
 local galaxies at fixed metallicity.
  In combination with simple photoionization models, the position of local and $z\sim2.3$ galaxies in excitation diagrams
 suggests that there is no significant change in the hardness of the ionizing spectrum at fixed metallicity from $z\sim0$ to $z\sim2.3$.
  We find that $z\sim2.3$ galaxies show no offset compared to low-metallicity local
 galaxies in emission line ratio diagrams involving only lines of hydrogen, oxygen, and sulfur, but show a systematic offset in diagrams involving
 [N~\textsc{ii}]$\lambda$6584.
  We conclude that the offset of $z\sim2.3$ galaxies from the local star-forming sequence in the [N~\textsc{ii}] BPT diagram
 is primarily driven by elevated N/O at fixed O/H compared to local galaxies.
  These results suggest that the local gas-phase and stellar metallicity sets the ionization state of star-forming regions
 at $z\sim0$ and $z\sim2$.
\end{abstract}

\keywords{galaxies: evolution --- galaxies: ISM --- galaxies: high-redshift}

\section{Introduction}\label{sec:intro}

The star-formation rate density of the universe peaked at $z\sim2$ \citep{hop06,mad14}.
  Understanding the physical conditions in star-forming regions during this epoch is essential for
 a complete description of the evolution of the stellar and gaseous content of galaxies.
  One method of probing these conditions is observing recently formed
 stars embedded in ionized gas clouds.  Rest-frame optical
 emission lines from HII regions encode a significant amount of information about the ionizing source and
 the physical conditions of the ionized gas.  A small set of physical properties appears to determine
 the strong optical emission line fluxes produced in HII regions, including the chemical abundance,
 the shape and normalization of the ionizing spectrum, the ionization state of the gas, and the gas
 density \citep{dop86,kew02,dop06a,dop06b}.

Over the past decade, a body of work has been produced showing that
 galaxies at $z\sim1-2$ display certain emission line ratios that are atypical of local star-forming galaxies
 \citep[e.g.,][]{sha05,erb06,liu08,hai09,hol14,nak14,ste14,sha15}.  These observations suggest that at least some of
 the physical conditions that influence emission line production in star-forming regions must be different in
 high-redshift galaxies.  It is well-established that galaxies at $z>1$ have lower metallicities than local
 galaxies at fixed stellar mass \citep{erb06,mai08,bel13,hen13,sto13,cul14,ste14,tro14,san15}.
  The electron density and ionization state of the gas also affect emission line production, but are less
 well-constrained and have fewer measurements at $z>1$.

Estimates of the electron density at $z\sim2$ have
 suggested that densities in high-redshift star-forming regions are significantly higher than what is typically observed locally
 \citep{hai09,leh09,bia10,shi14}.  
  However, the small and heterogeneous galaxy samples used in these studies leave the electron density of typical star-forming
 regions at $z\sim2$ poorly constrained.  Electron density estimates in a large, representative sample of $z\sim2$ galaxies are
 needed to obtain robust constraints on the typical electron densities at high redshift.

The typical ionization state of gas in $z\sim2$ star-forming regions appears to differ from that of local star-forming regions.
  Galaxies at $z>1$ display higher values of [O~\textsc{iii}]$\lambda$5007/H$\beta$ and
 [O~\textsc{iii}]$\lambda\lambda$4959,5007/[O~\textsc{ii}]$\lambda\lambda$3726,3729 than are typical of local galaxies, leading
 to the conclusion that high-redshift galaxies have higher ionization parameters than local galaxies, on average
 \citep{bri08,hai09,hol14,nak14,shi14}.  A systematic investigation of the variation of the ionization state with other
 galaxy properties at $z\sim2$ is necessary to uncover the cause of these elevated ionization parameters.

In this paper, we investigate the physical properties of star-forming regions at $z\sim2.3$, specifically the electron density
 and ionization state, using a large, systematically-selected sample from the MOSFIRE Deep Evolution Field (MOSDEF) survey.
  Until recently, samples of rest-frame optical spectra of $z\sim2$ galaxies were small, heterogeneous, and usually included
 only a subset of the strongest nebular emission lines.
  These samples reflected the difficulty of obtaining spectra of faint galaxies in the near-infrared,
 typically using long-slit spectrographs covering one near-infrared band at a time.
  With the development of sensitive near-infrared detectors and multi-object near-infrared spectrographs on 8-10~m class telescopes,
 large samples of $z\sim2$ galaxies with rest-frame optical emission line measurements across multiple near-infrared
 bands are being assembled for the first time \citep{ste14,kri15}.  These spectra contain a wealth of
 diagnostic information that probes the star-formation rate (SFR), dust attenuation, gas density,
 ionization state, chemical enrichment, and more for $z\sim2$ galaxies.
  Building on the work of \citet{sha15}, we utilize the full range of strong optical emission lines
 to investigate the physical properties of HII regions using various diagnostic line ratios.
  In combination with rich datasets
 at lower redshifts, such measurements make it possible to understand how conditions in star-forming regions
 have evolved over the past $\sim10$~Gyr of cosmic history.

Changes in the physical conditions of star-forming gas are thought to be the cause of the offset of $z>1$ galaxies from the
 local sequence of star-forming galaxies in the [O~\textsc{iii}]$\lambda5007$/H$\beta$ vs. [N~\textsc{ii}]$\lambda6584$/H$\alpha$
 excitation diagram \citep{sha05,liu08,hai09,ste14,sha15}.  It has been proposed that this offset is caused by higher gas density/pressure
 \citep{kew13t}, systematically harder ionizing spectra \citep{ste14}, higher ionization parameters \citep{bri08}, or
 an enhancement in the N/O ratio \citep{mas14,sha15} in high-redshift galaxies in comparison to what is typically
 observed in the local universe.  The offset could be caused by a combination of some or all of these parameters.
By characterizing the differences in the density and ionization state of $z\sim2$ and $z\sim0$ star-forming galaxies,
 as we do here, we can gain a better understanding of
 which parameters drive the offset in the [O~\textsc{iii}]$\lambda5007$/H$\beta$ vs. [N~\textsc{ii}]$\lambda6584$/H$\alpha$ excitation diagram,
 and the relative importance of each.  Determining the evolution of these properties with redshift also has implications
 for the applicability of local metallicity calibrations at $z\sim1-2$.

This paper is structured as follows.
  In Section~\ref{sec:obs}, we briefly describe the MOSDEF survey, along with the observations, reduction, and measurements.
  We estimate the typical electron density in $z\sim2.3$ star-forming regions and characterize the evolution of density
 with redshift in Section~\ref{sec:dens}.  In Section~\ref{sec:ionparam}, we investigate the ionization state of
 $z\sim2.3$ galaxies and its dependence on global galaxy properties and metallicity indicators.  We propose a scenario
 in which galaxies at $z\sim2.3$ have the same ionization parameter as galaxies at $z\sim0$ at fixed metallicity.  In
 Section~\ref{sec:discussion}, we
 provide evidence supporting our proposed scenario and discuss the implications for the interpretation of observed
 emission-line ratios of $z\sim2.3$ galaxies, including the offset in the [O~\textsc{iii}]$\lambda5007$/H$\beta$ vs.
 [N~\textsc{ii}]$\lambda6584$/H$\alpha$ diagram.  We conclude by summarizing our main results in Section~\ref{sec:summary}.

We adopt the following shorthand abbreviations to refer to commonly used emission line ratios:
\begin{equation}
\mbox{O}_{32}=[\mbox{O}~\textsc{iii}]\lambda\lambda4959,5007/[\mbox{O}~\textsc{ii}]\lambda\lambda3726,3729
\end{equation}
\begin{equation}
\mbox{R}_{23}=([\mbox{O}~\textsc{iii}]\lambda\lambda4959,5007 + [\mbox{O}~\textsc{ii}]\lambda\lambda3726,3729)/\mbox{H}\beta
\end{equation}
\begin{equation}
\mbox{O3N2}=([\mbox{O}~\textsc{iii}]\lambda5007/\mbox{H}\beta)/([\mbox{N}~\textsc{ii}]\lambda6584/\mbox{H}\alpha)
\end{equation}
\begin{equation}
\mbox{N2}=[\mbox{N}~\textsc{ii}]\lambda6584/\mbox{H}\alpha
\end{equation}
Throughout this paper, the term ``metallicity" is used synonymously with gas-phase oxygen abundance (O/H) unless otherwise mentioned.
  We adopt a $\Lambda$-CDM cosmology with $H_0=70$~km~s$^{-1}$~Mpc$^{-1}$, $\Omega_m=0.3$, and $\Omega_{\Lambda}=0.7$.

\section{Observations}\label{sec:obs}

We use data taken during the first two years (2012B-2014A) of the MOSFIRE Deep Evolution Field survey.
  We briefly describe the MOSDEF survey, observations, reduction, and derived quantities here.
  Full technical details of the survey strategy, observations, reduction pipeline, and sample
 characteristics can be found in \citet{kri15}.  We additionally use data from the
 Sloan Digital Sky Survey \citep[SDSS;][]{yor00} Data Release 7 \citep[DR7;][]{aba09} catalog
 to select local comparison samples for studying evolution with redshift.
  Emission-line measurements and galaxy properties are taken from the MPA-JHU catalog of measurements
 for SDSS DR7.\footnote[6]{Available at \texttt{http://www.mpa-garching.mpg.de/SDSS/DR7/}}

\subsection{The MOSDEF Survey}\label{sec:mosdef}

The MOSDEF survey is an ongoing multi-year project in which we are obtaining rest-frame optical spectra of galaxies
 at $z\sim1.4-3.8$ with the goal of transforming the understanding of the gaseous, stellar, dust,
 and black hole content of galaxies at that epoch in cosmic history.  This project utilizes the
 Multi-Object Spectrometer For Infra-Red Exploration \citep[MOSFIRE;][]{mcl12} on the 10 m Keck I
 telescope.  Potential for scientific gain from the MOSDEF dataset is maximized by targeting objects
 in the AEGIS, COSMOS, and GOODS-N extragalactic fields with extensive multi-wavelength
 ancillary data.  These data include \textit{Hubble Space Telescope} (\textit{HST}) imaging from
 the Cosmic Assembly Near-infrared Deep Extragalactic Legacy Survey
 \citep[CANDELS;][]{gro11,koe11} and grism spectroscopy from the 3D-HST survey \citep{bra12a}, as well as observations
 from {\it Chandra}, {\it Spitzer}, {\it Herschel}, VLA, and ground-based observatories in the optical
 and near-infrared.

In the MOSDEF survey, we target galaxies in the three redshift windows $1.37 \leq z \leq 1.70$, $2.09 \leq z \leq 2.61$,
 and $2.95 \leq z \leq 3.80$, where the redshift ranges are selected such that strong optical emission-line
 features fall within windows of atmospheric transmission in the Y, J, H, or K near-infrared bands.  This targeting strategy
 leads to coverage of [O\textsc{ii}]$\lambda\lambda$3726,3729, H$\beta$, and [O~\textsc{iii}]$\lambda\lambda$4959,5007
 for all three redshift bins, as well as [N~\textsc{ii}]$\lambda\lambda$6548,6584, H$\alpha$, and
 [S~\textsc{ii}]$\lambda\lambda$6716,6731 for the $z\sim1.5$ and $z\sim2.3$ bins.  These strong optical
 emission lines are powerful diagnostics for determining physical properties of galaxies including
 dust content \citep{red15}, star-formation rate (SFR) \citep{shi15}, chemical abundance \citep{san15}, gas density, ionization state, and black hole
 activity \citep{coi15}.  Targets are selected from the 3D-HST photometric and spectroscopic catalogs \citep{ske14} based on their rest-frame optical
 (observed H-band) magnitudes and redshifts (grism or spectroscopic redshifts, if available, and photometric
 redshifts otherwise).  Galaxies are targeted down to
 \textit{HST}/WFC3 F160W AB magnitudes of 24.0 at $z\sim1.5$, 24.5 at $z\sim2.3$, and 25.0 at $z\sim3.4$.  Targets
 with pre-existing spectroscopic redshifts, including grism redshifts or ground-based spectroscopic redshifts,
 are given higher priority, as are galaxies with brighter F160W magnitudes.  As described in \citet{kri15}, the MOSDEF
 survey will obtain spectra
 for $\sim1500$ galaxies when complete, with $\sim750$ at $z\sim2.3$, and $\sim400$ each at $z\sim1.5$ and $z\sim3.4$.

\subsection{Observations and Reduction}

We use data from the first two observing seasons of the MOSDEF survey.  Observations were taken on ten observing
 runs from 2012 December to 2014 May, during which 21 MOSFIRE masks were observed.  The first observing run, taking place
 in 2012 December, was a pilot program during which we observed one mask each in the GOODS-S and UDS CANDELS fields due to
 the limited visibility of the primary target fields.  This work focuses on the $z\sim2.3$ redshift bin, and, accordingly,
 we only describe the observations for this redshift interval.  MOSDEF targets at $z\sim2.3$ are observed in the J,
 H, and K near-infrared bands, with [O\textsc{ii}]$\lambda\lambda$3726,3729 in J, H$\beta$ and
 [O~\textsc{iii}]$\lambda\lambda$4959,5007 in H, and [N~\textsc{ii}]$\lambda\lambda$6548,6584, H$\alpha$, and
 [S~\textsc{ii}]$\lambda\lambda$6716,6731 in K.  Observed MOSFIRE masks each contain $\sim30$ slits
 with widths of 0\secpoint7, yielding a spectral resolution of 3300, 3650, and 3600 in J, H, and K bands, respectively.
  One slit on each mask was placed on a reference star used in the reduction.
  Masks were typically observed using an ABA'B' dither pattern, with individual exposure times at each dither position of
 120 seconds in J and H, and 180 seconds in K.  The total exposure time per filter per mask was typically 2 hours.

The raw data were reduced using a custom IDL pipeline developed by the MOSDEF team and described in detail
 in \citet{kri15}.
  The raw science frames were flatfielded and sky subtracted, cosmic rays were identified and masked, and
 the two-dimensional spectra were rectified.  Individual exposures were combined and the resulting
 spectrum was flux calibrated.  Shape correction due to varying spectral response
 with wavelength and telluric absorption features was achieved using observations of B8-A1 V standard stars
 matched to the typical air mass of science observations.  Flux calibration was performed by requiring the flux density of the reference
 star on a mask to match its cataloged broadband photometry.  For each slit, a two-dimensional error spectrum was
 produced accounting for Poisson counting uncertainty for the observed intensity per pixel and read noise.
  From the two-dimensional science and error spectra, one-dimensional spectra were produced using the optimal
 extraction technique.  Spectra for any detected objects serendipitously falling on the slit were also
 extracted (Freeman et al., in prep.).  The final flux calibration was achieved by applying a slit-loss correction term to the extracted
 science spectra on an individual basis.  The fraction of light from an object falling outside of the slit was
 estimated using a two-dimensional Gaussian fit to the F160W image of a galaxy convolved with the seeing
 estimate for each mask and filter.  The flux calibration was checked by verifying that objects with detected
 continuum had flux densities consistent with broadband photometry.

\subsection{Measurements and Derived Quantities}\label{sec:meas}

Emission line fluxes were measured by fitting Gaussian profiles to emission lines in the one-dimensional
 spectra, while the uncertainty on the emission line flux was based on the 68th percentile width of
 the distribution of measured fluxes obtained by perturbing the spectrum according to the error spectrum
 and refitting the emission line 1,000 times \citep{red15}.  All emission lines were fit with a single Gaussian profile except
 for the cases of [N~\textsc{ii}]$\lambda\lambda$6548,6584 and H$\alpha$, fit
 simultaneously with a triple Gaussian, and [O~\textsc{ii}]$\lambda\lambda$3726,3729, fit simultaneously
 with a double Gaussian and described in detail in Section~\ref{sec:densitysample}.
  The highest signal-to-noise emission line of each object, typically H$\alpha$ or [O~\textsc{iii}]$\lambda$5007,
 was used to constrain the centroid and width of the other emission lines and measure the redshift.
  Stellar masses were estimated by utilizing the stellar population synthesis models of \citet{con09} with the SED-fitting code
 FAST \citep{kri09} using the measured MOSDEF spectroscopic redshifts and broadband photometric catalogs
 assembled by the 3D-HST team \citep{ske14} spanning observed optical to mid-infrared.  The \citet{cal00}
 attenuation curve and a \citet{cha03} IMF were assumed.  Uncertainties on the stellar
 masses were estimated using a Monte Carlo method where the input photometry was perturbed according to
 the errors and the SED was refit 500 times \citep{kri15}.  SFRs were estimated from dust-corrected H$\alpha$ luminosities
 using the \citet{ken98} calibration converted to a \citet{cha03} IMF \citep{shi15}.  Balmer line fluxes were first
 corrected for stellar Balmer absorption using the best-fit SEDs \citep{red15}, and the dust correction was
 estimated using H$\alpha$/H$\beta$ assuming an intrinsic ratio of 2.86 and the \citet{car89} extinction
 curve.

Emission line fluxes are dust-corrected in line ratios involving lines significantly separated
 in wavelength (O$_{32}$ and R$_{23}$), but are not dust-corrected in line ratios featuring lines with a close
 wavelength spacing (O3N2, N2, [O~\textsc{iii}]/H$\beta$, and [S~\textsc{ii}]/H$\alpha$).
  Uncertainties on emission line ratios are estimated using a monte carlo method by perturbing each individual line flux according
 to its uncertainty, recalculating the emission line ratio, and repeating the preceding steps 10,000 times to build up a distribution
 of perturbed line ratios.  The uncertainty on the line ratio is determined from the 68th percentile width of this distribution.
  Error estimates on O$_{32}$ and R$_{23}$ values include uncertainty in the Balmer decrement by recalculating the
 dust correction for each realization.

After removing those objects identified as AGN based on their X-ray or IR properties \citep{coi15},
 there are 225 star-forming galaxies confirmed to be in the $z\sim2.3$ redshift interval spanning a stellar mass range of
 $10^{8.97}-10^{11.64}$~M$_{\odot}$ with a median stellar mass of $10^{9.99}$~M$_{\odot}$.
  The subset of these galaxies with H$\alpha$ and H$\beta$ detected (67\%) has SFRs spanning
 $1.61-323$~M$_{\odot}$/yr with a median SFR of 21.6~M$_{\odot}$/yr, and spans a range in stellar mass
 of $10^{8.97}-10^{11.22}$~M$_{\odot}$ with a median stellar mass of $10^{10.0}$~M$_{\odot}$.
  \citet{shi15} have shown that the SFR-\mstar\ relation of MOSDEF $z\sim2.3$ star-forming galaxies is similar to
 what is observed in other studies that employ different SFR indicators (see their Figure 8).
  Accordingly, the $z\sim2.3$ star-forming galaxy population recovered by the MOSDEF survey appears to be representative of the
 range of SFRs spanned by star-forming galaxies at these stellar masses and this redshift \citep{red12,whi14,shi15}.

\section{Electron Densities}\label{sec:dens}

The electron density of star-forming regions affects the fluxes of collisionally excited lines and is thus
 an important quantity to measure as an input parameter to photoionization models.\footnote[7]{The models described in
 Section~\ref{sec:discussion} actually use the hydrogen gas density as an input parameter.  In HII regions, the
 hydrogen gas is fully ionized such that the electron density is a good proxy for the hydrogen gas density.}
  The electron density can be estimated using the line fluxes of components of a doublet of a single
 species in which the two levels of the doublet have different collision strengths and radiative
 transition probabilities \citep{ost06}.  The flux observed in each component of the doublet is dependent on the
 relative population in each energy level, which is sensitive to the density of electrons available
 for collisional excitation and de-excitation.  Rest-frame optical spectra
 provide access to two strong emission line doublets useful for estimating the electron density, namely
 [O~\textsc{ii}]$\lambda\lambda3726,3729$ and [S~\textsc{ii}]$\lambda\lambda6716,6731$.  We use the ratios
 of the components of these doublets to estimate densities for local SDSS galaxies and high-redshift
 galaxies from the MOSDEF sample.

\subsection{Methods}\label{sec:methods}

We have written a python script that solves a 5-level atom approximation of the O~\textsc{ii} and
 S~\textsc{ii} ions for the relative populations in the second and third energy levels.
  Decays from these two energy levels produce the two components of the [O~\textsc{ii}]$\lambda\lambda3726,3729$
 and [S~\textsc{ii}]$\lambda\lambda6716,6731$ doublets.  We set up a
 detailed balance of transitions into and out of each of the five energy levels via radiative
 decay and collisional excitation and de-excitation, assuming the system is in thermal
 equilibrium.  The detailed balance provides a system of equations that can be solved for the relative
 populations at a given density.  Given $n_2$ and $n_3$, the relative populations in the second and third
 energy levels, the ratio of the line fluxes is given by
\begin{equation}
\frac{F_3}{F_2} =\frac{E_{31} n_3 A_{31}}{E_{21} n_2 A_{21}} \approx\frac{n_3 A_{31}}{n_2 A_{21}}
\end{equation}
where $F_i$ is the emission line intensity of decay from the $i$th level to the ground state, $E_{i1}$ is the
 energy difference between the $i$th level and the ground state, and $A_{i1}$ is the transition probability
 of the $i$th level to the ground state.

Calculating the proper emission line ratio corresponding to a given density requires accurate knowledge of the transition
 probabilities and collision strengths of each transition between the five energy levels.
  Recent investigations \citep{nic13,dop13} have suggested advantages in using the most up-to-date collision
 strength and transition
 probability atomic data instead of outdated values included in the IRAF routine \textit{temden}~\citep{temden}.
  Motivated by these studies, we adopt the effective collision strengths
 from \citet{tay07} for O~\textsc{ii} and \citet{tay10} for S~\textsc{ii}, while the transition probabilities
 for both species are taken from the NIST MCHF database \citep{nistmchf}.
  Using other estimates of the collision strengths and transition probabilities can change the calculated
 electron densities by up to $\sim30\%$.
  We verified that our script can exactly match IRAF~\textit{temden} when using the same atomic data.

  The effective
 collision strengths have some temperature dependence and have been calculated over a range of
 electron temperatures from 2000~K to 100,000~K for O~\textsc{ii}
 and 5000~K to 100,000~K for S~\textsc{ii}.  We adopt the effective collision strengths calculated with an
 electron temperature of 10,000~K, a representative equilibrium temperature of HII regions that are neither
 metal-rich nor metal-poor.  We note that the temperature dependence is not negligible.  Assuming an electron
 temperature of 7,000~K yields electron densities that are $\sim15-20\%$ lower at a fixed line ratio,
 while an electron temperature of 15,000~K gives densities that are higher by the same amount.  Because we assume
 a fixed electron temperature of 10,000~K, we are likely overestimating the electron density in metal-rich
 galaxies, and underestimating the density in galaxies that are metal-poor.  However, the uncertainty introduced by this
 assumption is smaller than the typical measurement uncertainty for individual $z\sim2.3$ galaxies.

\begin{figure}
 \centering
 \includegraphics[width=\columnwidth]{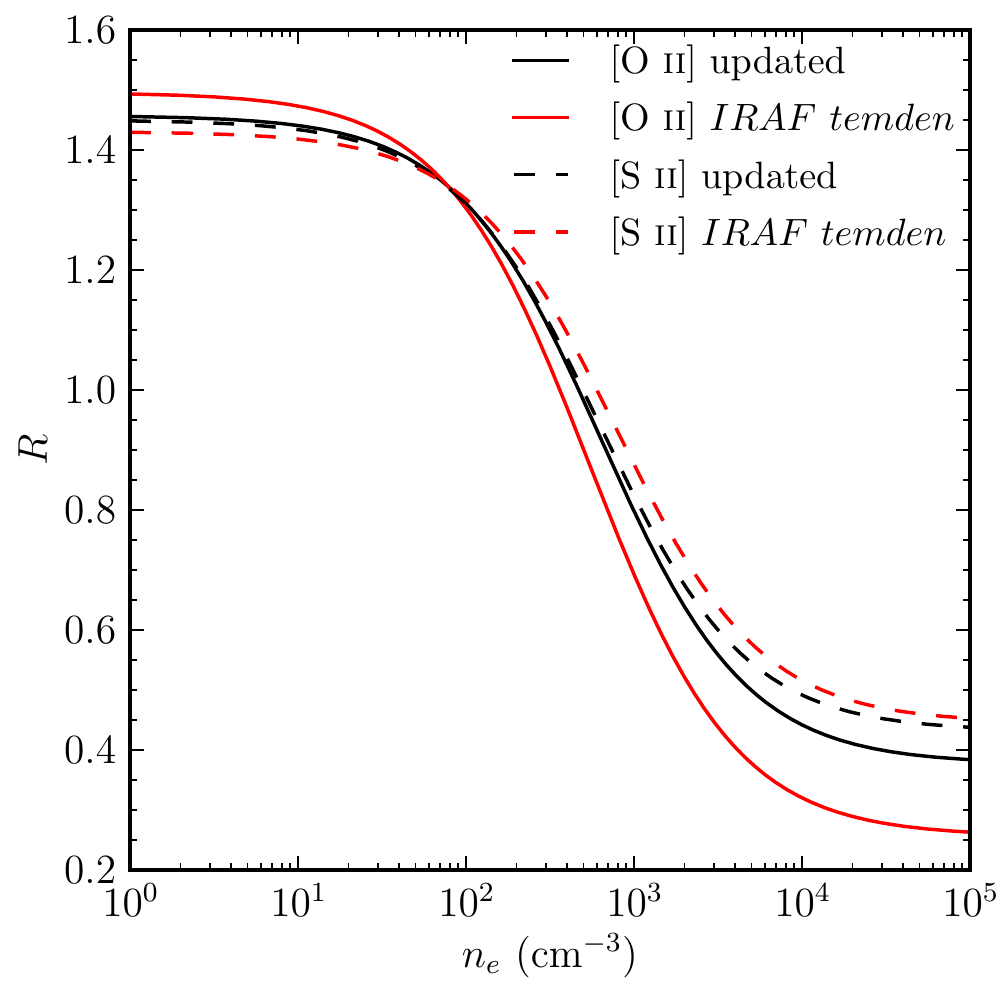}
 \caption{$R$ vs. $n_e$ curves (equation~\ref{eq:frat}) from IRAF temden (red) and our five-level atom python script using new atomic data (black), where
 $R=$[O~\textsc{ii}]$\lambda$3729/$\lambda$3726 (solid) or [S~\textsc{ii}]$\lambda$6716/$\lambda$6731 (dashed).
}\label{fig:newvsiraf}
\end{figure}

We calculate the line ratios [O~\textsc{ii}]$\lambda3729/\lambda3726$ and [S~\textsc{ii}]$\lambda6716/\lambda6731$
 over a range of electron densities of $\log{\left( \frac{\mbox{n}_e}{\mbox{cm}^{-3}}\right)}=0$ to 5 in 0.01 dex steps.
  The result is well fit by a function of the form
\begin{equation}\label{eq:frat}
R(n_e)=a \frac{b + n_e}{c + n_e}
\end{equation}
where $R$=[O~\textsc{ii}]$\lambda$3729/$\lambda$3726 or [S~\textsc{ii}]$\lambda$6716/$\lambda$6731 is the line flux ratio.  
 The best-fit parameters using up-to-date atomic data are shown in Table~\ref{tab:bestfit} for [O~\textsc{ii}] and
 [S~\textsc{ii}].  Figure~\ref{fig:newvsiraf}
 shows this diagnostic relation for the new atomic data (black) and the relation from \textit{IRAF temden}
 (red) for both [O~\textsc{ii}] (solid) and [S~\textsc{ii}] (dashed).  It can be seen in Figure~\ref{fig:newvsiraf} that
 the line ratio asymptotically approaches a theoretical maximum value in the low-density limit and a theoretical
 minimum value in the high-density limit.  We calculate the theoretical maximum and minimum line ratios at densities of 1~cm$^{-3}$
 and 100,000~cm$^{-3}$, respectively, and show these values in Table~\ref{tab:bestfit}.
  Inverting equation~\ref{eq:frat} yields the density as a function of the line ratio
\begin{equation}\label{eq:edens}
n_e(R)=\frac{cR - ab}{a - R}
\end{equation} 
which we use to calculate electron densities.  Uncertainties on individual density measurements are estimated by
 converting the upper and lower 68th percentile uncertainties on the line ratio into electron densities, where
 the upper (lower) uncertainty in line ratio corresponds to the lower (upper) uncertainty in density.

\begin{table}[t]
 \centering
 \caption{Coefficients and limiting line ratios for [O~\textsc{ii}] and [S~\textsc{ii}] in equations~\ref{eq:frat}~and~\ref{eq:edens}}\label{tab:bestfit}
 \begin{tabular}{ c | c c c | c c}
   \hline\hline
   $R$ & $a$ & $b$ & $c$ & $R_{\rm min}$\tablenotemark{a} & $R_{\rm max}$\tablenotemark{b} \\ \hline
   {[O~\textsc{ii}]$\lambda$3729/$\lambda$3726} & 0.3771 & 2,468 & 638.4 & 0.3839 & 1.4558 \\
   {[S~\textsc{ii}]$\lambda$6716/$\lambda$6731} & 0.4315 & 2,107 & 627.1 & 0.4375 & 1.4484 \\
   \hline
 \end{tabular}
 \tablenotetext{1}{Theoretical minimum line ratio calculated in the high-density limit of 100,000~cm$^{-3}$}
 \tablenotetext{2}{Theoretical maximum line ratio calculated in the low-density limit of 1~cm$^{-3}$}
\end{table}

\subsection{Sample}\label{sec:densitysample}

For objects with electron densities in the range $\sim100-1,000$~cm$^{-3}$, the line ratio will be fairly close to unity, as can
 be seen in Figure~\ref{fig:newvsiraf}.  In this regime, relatively small changes in the line flux can result in
 large changes to the inferred electron density.  Therefore, it is imperative that the doublets used to infer electron
 densities be free of any contamination from skylines or poor line profile fitting.  The [S~\textsc{ii}] doublet
 is well separated and the two components are fit separately with single Gaussian profiles.  The [O~\textsc{ii}]
 doublet is well-resolved but the two components are blended and must be fit with two Gaussian profiles simultaneously.
  When fitting the [O~\textsc{ii}] doublet, we constrain the separation of the centroids of the two components to be
 within $0.5$~\AA\ of the nominal separation of 2.78~\AA\ in the rest frame.  We also require the widths of the two
 components to match each other exactly and to be no more than 10\% larger than the velocity width inferred from
 the highest signal-to-noise line in the object's spectrum, typically H$\alpha$ or [O~\textsc{iii}]$\lambda$5007.
  This method yields robust fits to [O~\textsc{ii}] doublets.

\begin{figure}
 \centering
 \includegraphics[width=\columnwidth]{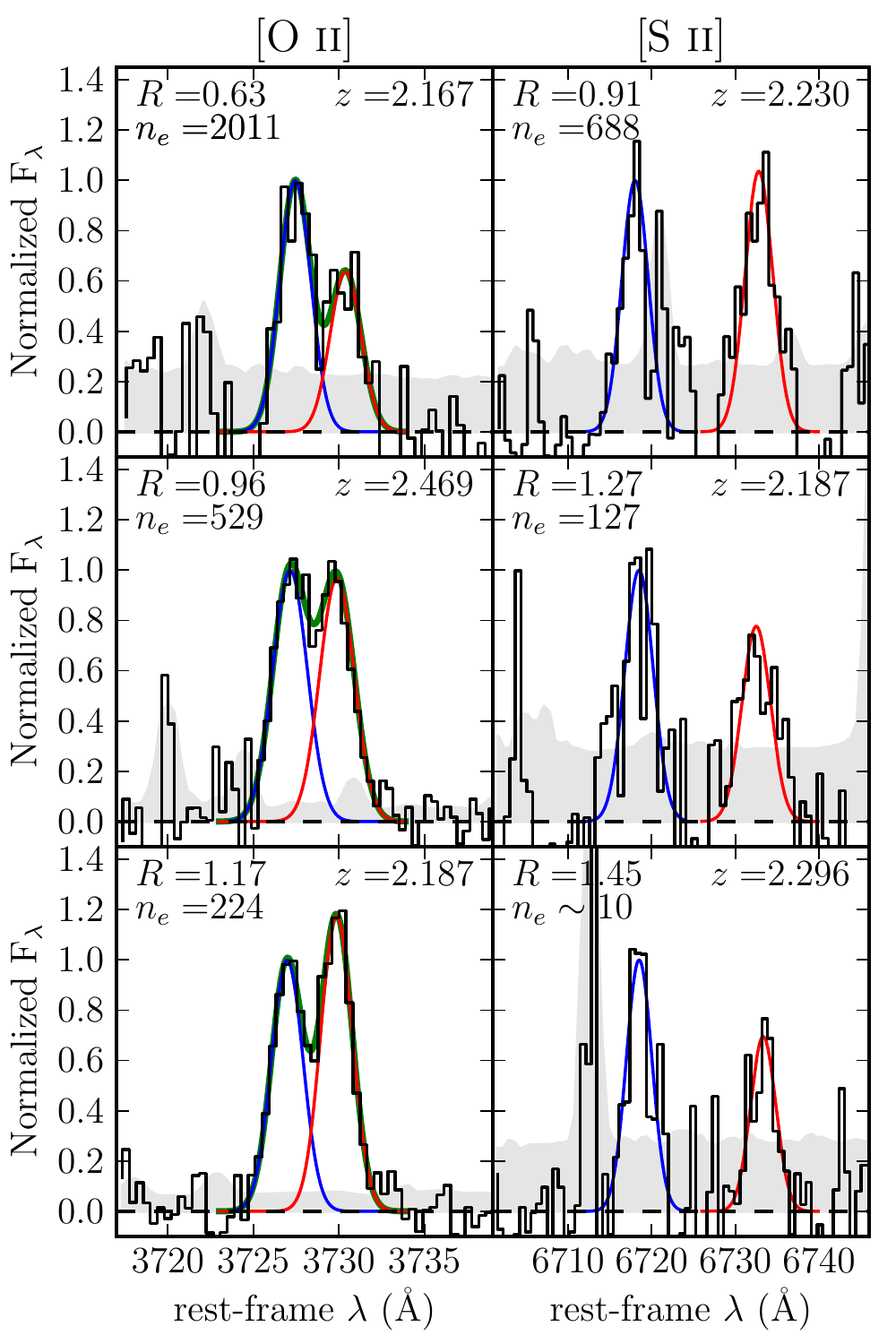}
 \caption{[O~\textsc{ii}]$\lambda\lambda$3726,3729 doublets (left column) and [S~\textsc{ii}]$\lambda\lambda$6716,6731
 doublets (right column) from six different objects over a range of line ratios and densities.
  The black line shows the continuum subtracted spectrum for each object.  The light gray band
 indicates the error spectrum for each object, while the blue and red lines show the Gaussian profile fits to the blueward
 and redward component of each doublet, respectively.  The green line shows the total [O~\textsc{ii}] profile.  In each panel,
 the spectrum has been normalized so that the blue component has a peak height of unity.
  The line ratio $R$, density ($n_e$) in cm$^{-3}$, and redshift is displayed for each
 object, with $R=$[O~\textsc{ii}]$\lambda$3729/$\lambda$3726 in the left column and
 $R=$[S~\textsc{ii}]$\lambda$6716/$\lambda$6731 in the right column.
}\label{fig:doublets}
\end{figure}

We select MOSDEF galaxies in the redshift range $2.0<z<2.6$ with S/N$\geq$3 in [O~\textsc{ii}]$\lambda\lambda$3726,3729
 or [S~\textsc{ii}]$\lambda\lambda$6716,6731 which have not been flagged as AGN based on their IR and X-ray properties
 \citep{coi15}.  This gives a sample of 97 [O~\textsc{ii}] doublets and 36 [S~\textsc{ii}] doublets at $z\sim2.3$.
  We visually inspected each of these doublets and removed those with significant skyline contamination or
 spurious detections.  One additional [S~\textsc{ii}] object with a very high value of
 $\log{(\mbox{[N\textsc{ii}}\lambda6584/\mbox{H}\alpha)}=-0.11$ indicating a probable AGN was also removed, giving a
 final density sample of 43 [O~\textsc{ii}] doublets and 26 [S~\textsc{ii}] doublets
 from 61 different targets at $\langle z\rangle = 2.24 \pm 0.12$.  This sample spans a range of stellar mass from
 10$^{8.97}-10^{11.22}$~M$_{\odot}$ with a median stellar mass of 10$^{10.10}$~M$_{\odot}$.
  Of the 61 galaxies, 8 (13\%) do not have measured SFRs, of which 3 galaxies do not have wavelength coverage of H$\beta$,
 2 galaxies do not have wavelength coverage of H$\alpha$, and the remaining 3 galaxies show significant skyline contamination
 of the H$\beta$ line.
  The other 53 objects have SFRs spanning $4.65-228$~M$_{\odot}$/yr with a median SFR of 29.7~M$_{\odot}$/yr.
  The density sample has only slightly higher median \mstar\ and SFR than the parent sample of MOSDEF $z\sim2.3$ star-forming
 galaxies (Section~\ref{sec:meas}) and is still representative of the fairly massive $z\sim2.3$ star-forming galaxy population.
  The 8 galaxies without SFR measurements are included in all parts of the analysis for which SFR is not required.
  Examples of [O~\textsc{ii}] and [S~\textsc{ii}]
 doublets from six different objects in the MOSDEF $z\sim2.3$ sample are shown in Figure~\ref{fig:doublets}, along with fits and inferred
 line ratios.  

We use a local comparison sample from SDSS to investigate evolution in the typical densities of star-forming
 regions.  We select galaxies from SDSS in the redshift range $0.04<z<0.1$ to attain a sample that
 is relatively free of aperture effects and limited to the local universe.  We require galaxies to have
 S/N$\geq$3 in H$\beta$, [O~\textsc{iii}]$\lambda$5007, H$\alpha$, [N~\textsc{ii}]$\lambda$6585,
 [S~\textsc{ii}]$\lambda$6716, and [S~\textsc{ii}]$\lambda$6731.  Detections in the first four of these lines are
 required to reject AGN using the empirical demarcation of \citet{kau03b}.  We do not require detection of the
 [O~\textsc{ii}] doublet.  The SDSS spectra have a spectral resolution
 of $\sim2000$, corresponding to a resolution element of $\sim1.9$~\AA~at 3727~\AA.  This resolution is
 insufficient to properly resolve and sample the [O~\textsc{ii}]$\lambda\lambda$3726,3729 doublet separated
 by 2.78~\AA\ (see Section~\ref{sec:res}).  Therefore, we only use the [S~\textsc{ii}] doublet to probe the electron density in the local
 comparison sample.  The local comparison sample contains 99,291 galaxies with $\langle z\rangle=0.0678$.

\subsection{The consistency of \textsc{[O~ii]} and \textsc{[S~ii]} electron densities}\label{sec:res}

Since our high-redshift electron density sample has a mixture of [O~\textsc{ii}] and [S~\textsc{ii}] doublet
 measurements while the local comparison sample only has reliable [S~\textsc{ii}] measurements, it is a
 useful exercise to evaluate the consistency of densities determined using these two ionic species to see
 if they can be directly compared.  To this end, we have assembled a sample of local HII regions from the
 literature with high-resolution, high-signal-to-noise spectroscopic observations with sufficient wavelength
 coverage to span [O~\textsc{ii}]$\lambda\lambda$3726,3729 and [S~\textsc{ii}]$\lambda\lambda$6716,6731.
  We performed a literature search and identified 32 galactic and extragalactic HII regions observed at high spectral resolution
 ($R\sim8,000-23,000$) with detections of both the [O~\textsc{ii}] and [S~\textsc{ii}] doublets
 \citep{gar05,gar06,gar07,lop07,est09,est13,est14}.

\begin{figure}
 \centering
 \includegraphics[width=\columnwidth]{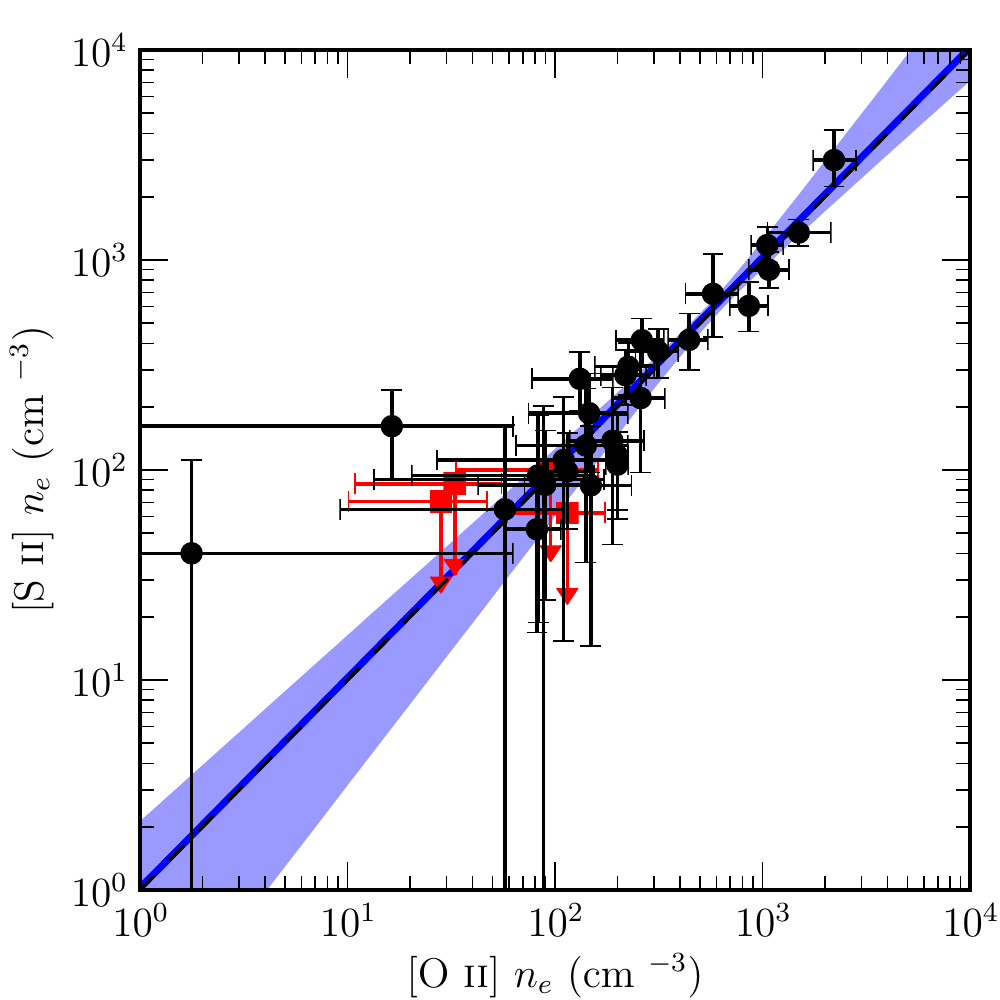}
 \caption{Comparison of density estimates from the [O~\textsc{ii}] and [S~\textsc{ii}] doublets for a sample
 of local HII regions with high-S/N, high-resolution spectra.  Black points denote density measurements for individual
 HII regions.  The four red squares show limits plotted at the upper 1$\sigma$ uncertainty bound on the [S~\textsc{ii}]
 density for objects that have higher [S~\textsc{ii}]$\lambda$6716/$\lambda$6731 than the maximum theoretically allowed
 value.  The black dashed line shows a one-to-one relationship.  The blue line and shaded blue region show the best-fit
 line and 1$\sigma$ confidence interval, respectively.  Parameters of the best-fit line are shown in equation~\ref{eq:bestfit}.
}\label{fig:siivsoiidensity}
\end{figure}

Electron densities and uncertainties are calculated with the same method outlined above
 using the published line fluxes and errors.  Densities and uncertainties are presented in
 Figure~\ref{fig:siivsoiidensity}.  Four of
 the 32 individual measurements have [S~\textsc{ii}]$\lambda$6716/$\lambda$6731 ratios that are higher than
 the theoretically allowed maximum in the low-density limit, and thus cannot be assigned a density.
  While these four objects have [S~\textsc{ii}]$\lambda\lambda$6716,6731 detected at greater than 3$\sigma$
 indicating they are in the low-density limit, we plot them as upper limits (red squares) where the data point is
 plotted at the [S~\textsc{ii}] density corresponding to the lower 1$\sigma$ uncertainty on the
 [S~\textsc{ii}] line flux ratio.  Fitting a line in logarithmic space yields the relationship
\begin{equation}\label{eq:bestfit}
\log\left(\frac{[\textsc{S~ii}]\ n_e}{\mbox{cm}^{-3}}\right) = 1.00^{+0.29}_{-0.12}\times \log\left(\frac{[\textsc{O~ii}]\ n_e}{\mbox{cm}^{-3}}\right) - 0.01^{+0.32}_{-0.80}
\end{equation}
where the 68th percentile confidence intervals are determined by perturbing the data points according to their
 uncertainties and refitting.  This best-fit line is shown in Figure~\ref{fig:siivsoiidensity}
 as a solid blue line while the light blue shaded region shows
 the 68th percentile confidence region around the best-fit line.  The relation between electron densities
 determined by [O~\textsc{ii}] and [S~\textsc{ii}] is completely consistent with a one-to-one relation (dashed black line).
  All four objects plotted as upper limits are also consistent with a one-to-one relation within the 1$\sigma$ uncertainties.
  There are 8 objects in the $z\sim2.3$ density sample that have density estimates from both the [O~\textsc{ii}] and
 [S~\textsc{ii}] doublets, but the number of galaxies is too small and measurement uncertainties are too large to
 perform a similar investigation at $z\sim2.3$.
  We assume that the relationship between densities of star-forming regions determined using [S~\textsc{ii}] and [O~\textsc{ii}]
 does not change with redshift.
  Thus, we conclude that densities determined from either ionic species in the $z\sim2.3$ sample can be directly compared with
 each other and with SDSS density measurements from [S~\textsc{ii}] doublets.

For this test, we specifically selected a sample with high S/N and very high spectral resolution so that each component of the
 doublets was well-detected and the [O~\textsc{ii}] doublet was well resolved.  We note that repeating
 this exercise with medium-resolution spectra ($R\sim1,000-2,000$) of local HII regions \citep{pei12,gar14,ber15}
 yields a relation in which [O~\textsc{ii}] electron densities are systematically overestimated with respect to
 [S~\textsc{ii}] electron densities.  This effect is very similar to what is seen in the SDSS sample ($R\sim2,000$).
  In order to have at least two resolution elements to sample the separation of the [O~\textsc{ii}] doublet
 components, a spectral resolution of $\Delta\lambda=1.39$~\AA\ at $\lambda=3727$~\AA\ is needed, corresponding
 to $R\sim2,700$.  MOSDEF observations adequately sample the [O~\textsc{ii}] doublet
 at $z\sim2.3$ with $R\sim3,300$ in the J band.

\subsection{Typical electron density at $z\sim2.3$ and $z\sim0$}

We would like to characterize the typical electron density in star-forming regions of $z\sim2.3$ galaxies
 and $z\sim0$ galaxies.  However, given the shape of the function in equation~\ref{eq:edens} and
 Figure~\ref{fig:newvsiraf}, a fairly
 symmetric distribution of line ratios leads to a very asymmetric distribution in electron densities.
  Furthermore, the diagnostic curve translating from line ratio to electron density is insensitive to
 the electron density at very low and very high densities, asymptotically approaching the theoretical maximum and minimum line
 ratio, respectively, in those two regimes.  Measured line ratios that fall outside of the theoretically allowed
 region due to measurement uncertainty can only be assigned limits in the low- or high-density extremes.
  We consider the low-density limit to refer to a density below $\sim10$~cm$^{-3}$ and the high-density
 limit to denote density above $\sim10,000$~cm$^{-3}$.
  For these reasons, we perform statistics on the line ratio distributions for each sample rather than the
 electron density distributions and infer typical electron densities based on the statistical properties
 of the line ratio distribution.

\begin{figure}
 \centering
 \includegraphics[width=\columnwidth]{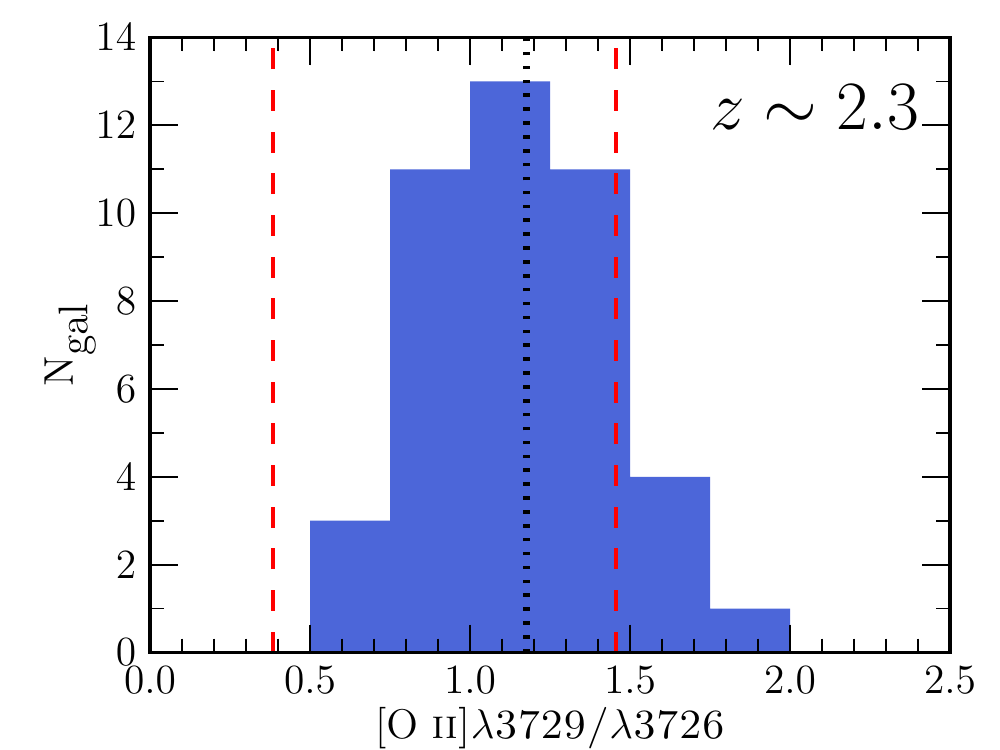}
 \includegraphics[width=\columnwidth]{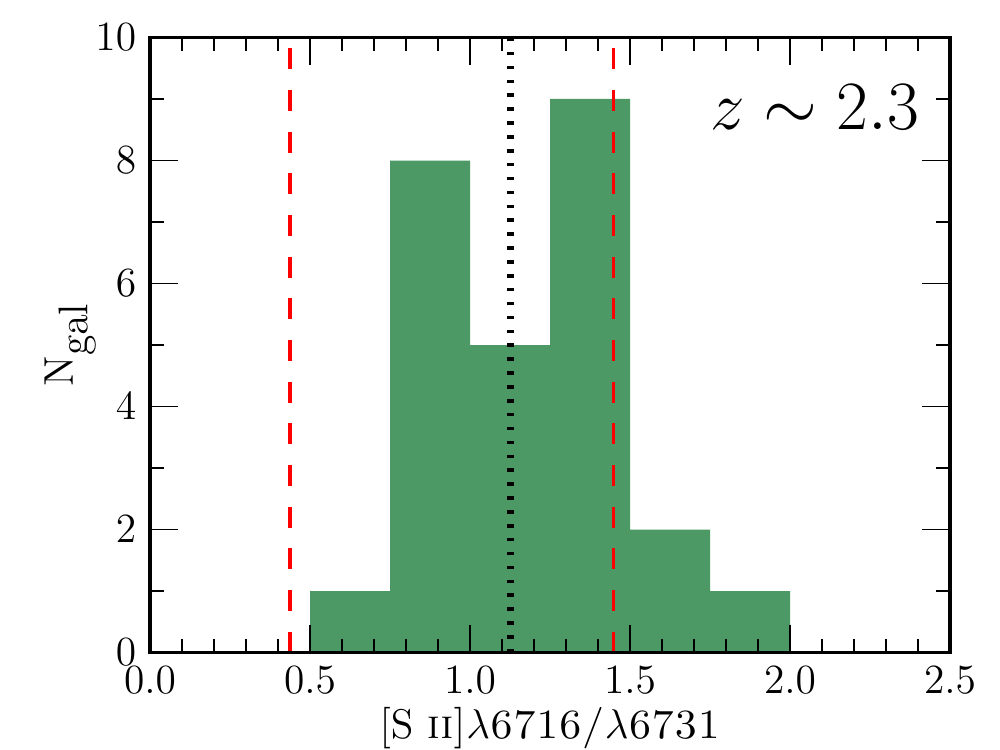}
 \caption{[O~\textsc{ii}]$\lambda3729/\lambda3726$ (top) and [S~\textsc{ii}]$\lambda6716/\lambda6731$ (bottom)
 line ratio distributions for 43 and 26 $z\sim2.3$ star-forming galaxies, respectively.  In each panel, the dotted black line shows the median
 line ratio (corresponding to an electron density of 225~cm$^{-3}$ for [O~\textsc{ii}] and 290~cm$^{-3}$ for [S~\textsc{ii}]),
 while the dashed red lines show the minimum and maximum theoretically-allowed line ratios from
 Table~\ref{tab:bestfit}.
}\label{fig:mosdefratiodist}
\end{figure}

\begin{figure}
 \centering
 \includegraphics[width=\columnwidth]{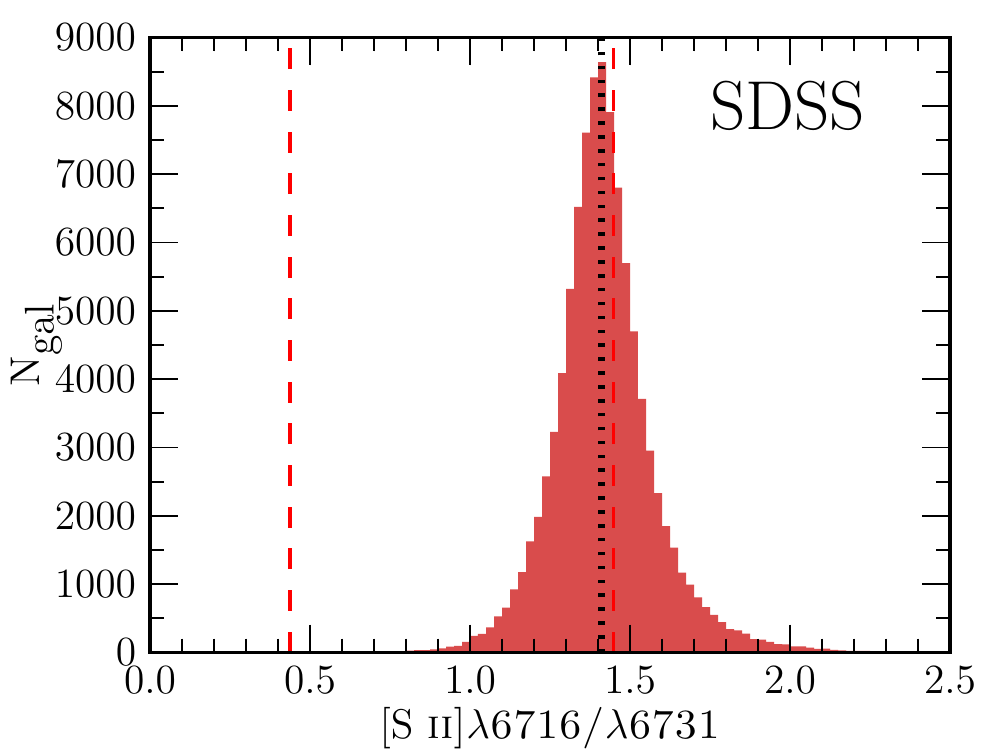}
 \caption{[S~\textsc{ii}]$\lambda6716/\lambda6731$ line ratio distribution for local star-forming
 galaxies from SDSS, with lines as in Figure~\ref{fig:mosdefratiodist}.  The median line ratio for local
 galaxies (corresponding to an electron density of 26~cm$^{-3}$) falls near the low-density limit.
}\label{fig:sdssratiodist}
\end{figure}

The distributions of line ratios of the $z\sim2.3$ star-forming galaxies are shown in the top and bottom panels
 of Figure~\ref{fig:mosdefratiodist} for the [O~\textsc{ii}] and [S~\textsc{ii}] doublets, respectively.
  We determine the typical electron density from the median line ratio of a given sample.  The uncertainty
 on the median is calculated using a bootstrap technique in which we randomly resample with replacement,
 perturb the emission line fluxes according to their uncertainties and recalculate the line ratio for each
 object in the new sample, take the median of the new perturbed sample, and repeat the preceding steps 1,000
 times to build up a well-sampled distribution of median values.  The reported lower and upper uncertainties  on the
 median are determined to be the 15.8-percentile and 84.2-percentile values, respectively, of the cumulative
 distribution function of the median.
  At $z\sim2.3$, we find a median [O~\textsc{ii}]$\lambda3729/\lambda3726$ ratio of 1.18$_{-0.10}^{+0.01}$ corresponding to
 an electron density of 225$_{-4}^{+119}$~cm$^{-3}$.  We find a median [S~\textsc{ii}]$\lambda6716/\lambda6731$ value of
 1.13$_{-0.06}^{+0.16}$ which gives an electron density of 290$_{-169}^{+88}$~cm$^{-3}$, consistent within the uncertainties
 with the density determined using [O~\textsc{ii}].  We measure a range of individual electron densities from the low-density limit
 to 2,500~cm$^{-3}$ and find that
 $z\sim2.3$ star-forming regions have a typical electron density of $\sim250$ cm$^{-3}$.

The typical density that we infer for high-redshift galaxies ($\sim250$~cm$^{-3}$) is in excellent
 agreement with what \citet{ste14} observed by stacking J-band observations of 113 galaxies at $z\sim2.3$,
 finding an average [O~\textsc{ii}]$\lambda3729/\lambda3726$ ratio of 1.16, corresponding to a density of
 243~cm$^{-3}$ using the atomic data adopted in this paper.  \citet{shim15} found a median
 electron density of 291~cm$^{-3}$ among 14 H$\alpha$ emitters at $z\sim2.5$ using the [O~\textsc{ii}] doublet.
  Previous observations of individual gravitationally lensed galaxies at $z\sim2$ suggested electron densities of
 $\sim1000$~cm$^{-3}$ \citep{hai09,bia10}, somewhat higher than the value we infer.  \citet{leh09} presented electron
 densities in the range $400-1200$~cm$^{-3}$ for 4 galaxies at $z\sim2.3$ in the SINS survey \citep{for09},
 also higher than our sample median. 
  As noted in Section~\ref{sec:methods}, different choices of atomic data can change the inferred densities by $\sim30\%$ and
 can lead to differences of this magnitude in measured densities reported by different authors.
  In comparison to these previous estimates, except for that of \citet{ste14}, our sample is larger and selected in
 a more systematic way, and the galaxies in our sample display properties representative of the SFR-\mstar\ relation
 at $z\sim2$, as shown in Section~\ref{sec:meas}.  As such, the density estimate presented here should hold true for a population of typical
 star-forming galaxies at $z\sim2.3$ with M$_*\gtrsim10^{9.5}$~M$_{\odot}$.

For the local comparison sample, we find SDSS star-forming galaxies have a distribution with a median
 [S~\textsc{ii}]$\lambda6716/\lambda6731$ ratio of 1.41, shown in Figure~\ref{fig:sdssratiodist},
 corresponding to an electron density of 26~cm$^{-3}$.
  The uncertainty on the SDSS median line ratio is less than 0.04\% due to the large number of
 galaxies in the sample.  We find that the typical electron density in star-forming regions increases
 by a factor of 10 from $z\sim0$ to $z\sim2.3$.
  The local median [S~\textsc{ii}]$\lambda6716/\lambda6731$ ratio is close to the theoretical maximum ratio of 1.4484,
 suggesting that local star-forming galaxies typically fall close to the low-density
 limit.  From the slope of the function in Figure~\ref{fig:newvsiraf}, it is apparent that the line ratios of
 both [O~\textsc{ii}] and [S~\textsc{ii}] are almost completely insensitive to the density when the electron
 density is below $~\sim10$~cm$^{-3}$, and are only mildly sensitive to the density below $~\sim50$~cm$^{-3}$.
  Even if we assume a conservative upper limit of 50~cm$^{-3}$ for the typical
 local density, we still observe a significant increase in electron density from $z\sim0$ to $z\sim2.3$.

The measurements above suggest that $z\sim2.3$ star-forming regions are typically denser than local
 star-forming regions by an order of magnitude.  We perform some tests to investigate the
 significance of the observed evolution in electron density.  First, note that the SDSS line ratio
 distribution shown in Figure~\ref{fig:sdssratiodist} is well-sampled and fairly narrow.  We find that
 89\% of the SDSS sample has higher [S~\textsc{ii}] ratios (lower electron densities) than the median
 $z\sim2.3$ [S~\textsc{ii}] ratio, while 64\% of the $z\sim2.3$ sample has lower [S~\textsc{ii}] ratios
 (higher electron densities) than the SDSS sample median.
  A two-sided Kolmogorov-Smirnov test on the $z\sim0$ and $z\sim2.3$
 [S~\textsc{ii}] distributions yields a K-S statistic of 0.452 and a p-value (the probability that the
 two samples were drawn from the same underlying distribution) of $7.63\times 10^{-6}$,
 indicating that the $z\sim0$ and $z\sim2.3$ density distributions are significantly different.

We are attempting to estimate the electron density in HII regions, but [S~\textsc{ii}] emission coming from other components
 of the ISM could contaminate measured [S~\textsc{ii}] ratios from integrated-light galaxy spectra.  
  The S~\textsc{ii} ion has an ionization energy of 10.36~eV, somewhat lower than that of hydrogen.
  Because of its lower ionization energy, the S~\textsc{ii} zone can extend beyond the boundary of an HII region.  Additionally,
 [S~\textsc{ii}]$\lambda\lambda$6716,6731 emission can be produced in diffuse ionized gas that is shock excited \citep{rey85,mar97}.
  To investigate the effects of contamination from a diffuse ionized ISM component, we used
 measurements of [S~\textsc{ii}] ratios of 44 HII regions in the star-forming spiral galaxy NGC~628 observed as part
 of the CHAOS survey \citep{ber15}.
  The spectra of these HII regions were attained by placing slits on top of individual HII regions and should contain
 very little light from the diffuse ISM.  We find that the median [S~\textsc{ii}]$\lambda$6716/$\lambda$6731 ratio of
 these HII regions is 1.39, corresponding to an electron density of 38~cm$^{-3}$.
  This ratio is nearly equivalent to the median [S~\textsc{ii}] ratio of 1.41 for the SDSS sample,
 suggesting that our estimate of the typical local HII region density from SDSS
 is not significantly biased by emission from diffuse ionized gas.  However, all of these HII regions are from a single
 galaxy, and their median density may not be representative of the entire local HII region population.
  Currently, there is insufficient knowledge of the ISM structure of
 $z\sim2$ galaxies to determine whether integrated-light spectra are significantly contaminated by emission from a diffuse
 component at that redshift.

\subsection{Electron density vs. galaxy properties}\label{sec:densgalprop}

\begin{figure*}
 \centering
 \includegraphics[width=\textwidth]{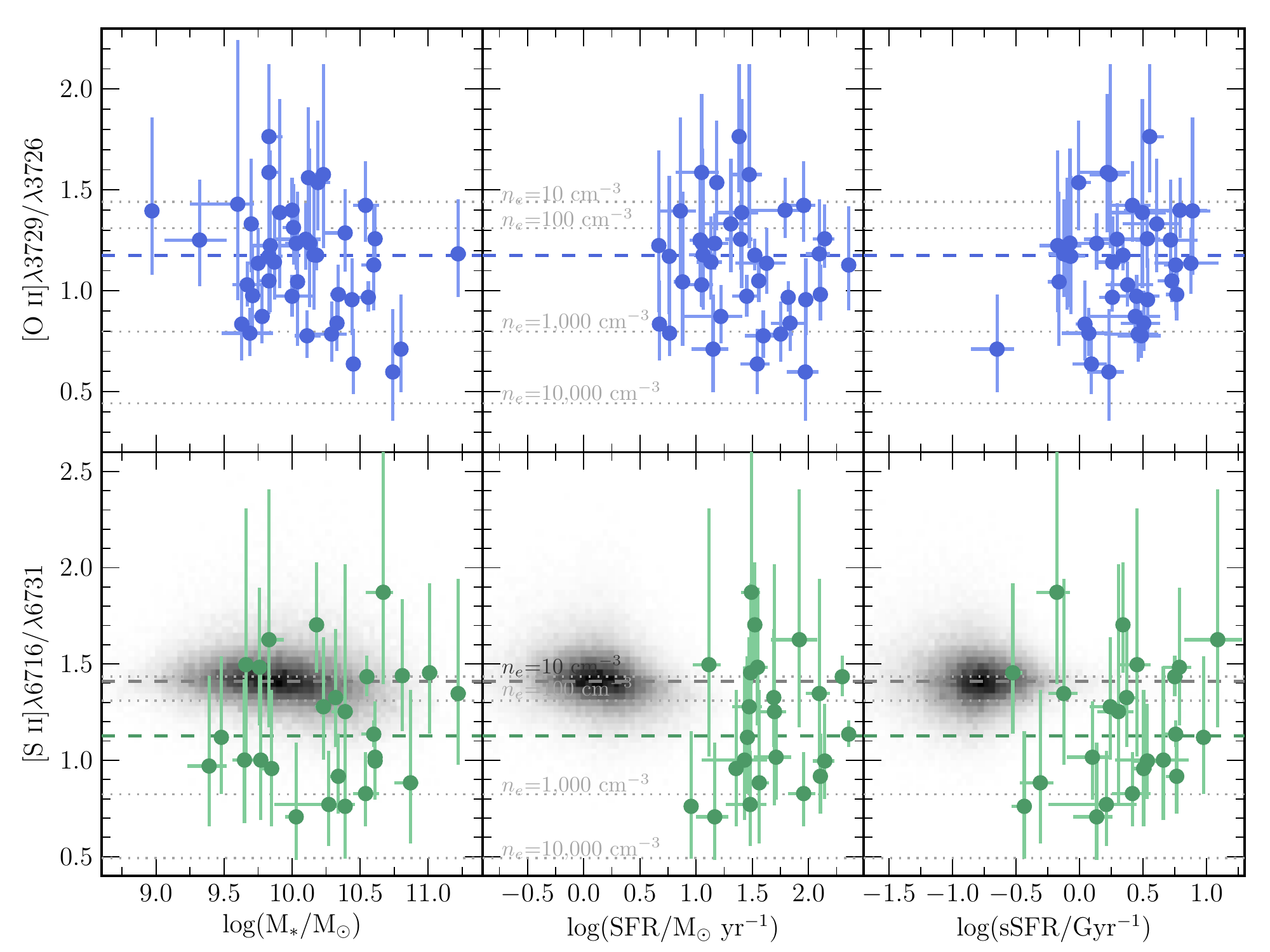}
 \caption{[S~\textsc{ii}]$\lambda\lambda$6716,6731 (bottom row) and [O~\textsc{ii}]$\lambda\lambda$3726,3729
 (top row) as a function of stellar mass (left), SFR (middle), and sSFR (right).  Blue and green points show
 the $z\sim2.3$ [O~\textsc{ii}] and [S~\textsc{ii}] density samples, respectively.  The gray two-dimensional
 histogram in the bottom row shows the distribution of the local comparison sample.  Spectra from SDSS do not
 have a high enough spectral resolution to resolve the components of the [O~\textsc{ii}] doublet.  The blue, green,
 and gray dashed lines show the median line ratios for the corresponding sample.  Dotted lines show the line
 ratios corresponding to denities of 10, 100, 1,000, and 10,000 cm$^{-3}$.
}\label{fig:densityvsgalaxyproperties}
\end{figure*}

We investigate whether the density of star-forming regions varies with other galaxy properties.
  As stated previously, the nature of the function converting between line ratio and electron density
 makes it difficult to work with distributions in density space, especially when some objects have
 measured line ratios that are outside of the theoretically allowed values.  For this reason we will
 look for relationships between density and galaxy properties using the line ratio as a proxy for the
 density.  We plot the line ratios against stellar mass (\mstar), star-formation rate (SFR),
 and specific star-formation rate (sSFR; SFR/\mstar) in Figure~\ref{fig:densityvsgalaxyproperties}.
  The middle and right panels of Figure~\ref{fig:densityvsgalaxyproperties} only include the subset
 of the $z\sim2.3$ density sample with H$\alpha$ and H$\beta$ detections.
  We note that the $z\sim2.3$ sample has significant overlap with the SDSS sample in \mstar, but
 the two are almost completely disjoint in SFR and sSFR.  This difference is consistent with the evolution of
 the SFR-\mstar\ relation with redshift \citep{shi15}.
  We do not see evidence for any significant trends in line ratio (electron density) as a function
 of stellar mass, SFR, or
 sSFR among the local SDSS sample or the MOSDEF $z\sim2.3$ sample.  This observation is confirmed by
 performing a Spearman correlation test on each sample in each parameter space.  No correlations
 are more significant than $\sim1\sigma$.

This result is in conflict with the recent work of \citet{shim15}, who found a correlation between electron
 density and sSFR at the 4$\sigma$ level for 14 $z\sim2.5$ H$\alpha$ emitters, and observed this correlation
 when stacking the spectra in two density bins.
  We do not see evidence
 for correlation among these properties with a larger and more representative sample that is generally consistent with
 that of \citet{shim15} in stellar mass and SFR.
  \citet{shim15} found
 no correlation between electron density and stellar mass, in agreement with our results.

\section{Ionization Parameter}\label{sec:ionparam}

The ionization state of ionized gas in star-forming regions refers to the interaction between the ionizing source
 and ionized gas. This interaction modulates the relative populations of different ionic species, which directly
 influence the observed emission line ratios. Multiple lines of evidence suggest that the ionization state in
 high-redshift star-forming regions may be systematically different from what is typically observed in the local
 universe in galaxies of similar masses \citep[e.g.,][]{hol14,nak14,ste14}.  The ionization parameter
 is useful for quantifying the ionization state, and encodes information about both the ionizing source and
 the surrounding ionized gas.  In this section, we will use emission line measurements of $z\sim2.3$ galaxies
 to investigate the relationship between ionization parameter and other galaxy properties in order to probe
 the ionization state of gas in $z\sim2$ star-forming regions.

\subsection{Definition of the ionization parameter}\label{sec:ionpdef}

We begin with some useful definitions related to the ionization state.
  Ionic species in HII regions are in ionization equlibrium, where the rate of ionization is equal to the
 rate of recombination.  The ionization equilibrium condition for ionic species $i$ can be written as
\begin{equation}
n_i \frac{Q_i}{4\pi r^2} \bar\sigma_{i} = n_i^+ n_e \alpha_i
\end{equation}
where $n_i$ is the number density of the ionic species, $Q_i$ is the rate of production of photons that can ionize
 species $i$, $\bar\sigma_{i}$ is the effective ionization cross-section, $n_i^+$ is the number density of the once-ionized state of $n_i$,
 $n_e$ is the electron density, and $\alpha_i$ is the recombination coefficient.  If the ionization energy of $n_i$ is close
 to that of hydrogen, which is the case for many of the ions that produce strong optical emission lines, then we can rearrange
 this expression and approximate the ratio of the relative populations in the higher and lower ionized states.
\begin{equation}\label{eq:ionfrac}
\frac{n_i^+}{n_i}\approx \frac{\bar\sigma_{i}}{\alpha_i} \frac{Q_0}{4\pi r^2 n_e}
\end{equation}
Here, $Q_0$ is the rate of production of hydrogen-ionizing photons ($h \nu \geq 13.6$~eV).  The first term ($\bar\sigma_{i}/\alpha_i$) on the right hand side
 of this expression is a constant that will change for each ionic species.  The second term ($Q_0/4\pi r^2 n_e$) contains only properties of the
 ionizing source and the gas, and is not dependent on the specific ionic species.  The dimensionless ionization parameter
 $\mathcal{U}$ is defined as the second
 term on the right-hand side of equation~\ref{eq:ionfrac} divided by the speed of light, $c$.
\begin{equation}
\mathcal{U}=\frac{Q_0}{4\pi r^2 c n_e}
\end{equation}
Accordingly, the ratio of the relative population in an upper ionization state to that in a lower ionization state scales directly with the
 ionization parameter.
  Since the electron density is approximately the hydrogen gas density in a fully-ionized plasma, the ionization parameter can
 be thought of as the ratio of the number density of hydrogen-ionizing photons to the number density of the hydrogen gas.
When working with
 HII regions it is convenient to define the dimensionless ionization parameter using the radius of a canonical
 Str\"omgren sphere, $R_S$, as the distance between the gas and the ionizing source.
\begin{equation}
\mathcal{U}=\frac{Q_0}{4\pi R_S^2 c n_e}
\end{equation}
Often, the dimensional ionization parameter, $q=c\times \mathcal{U}$, is used instead, which is the ratio of the flux of
 ionizing photons at the Str\"omgren radius to the hydrogen number density.
  The definition of the Str\"omgren radius, based on a balance between ioniziation and recombination rates
 assuming case B recombination, is
\begin{equation}
R_S=\left( \frac{3 Q_0}{4 \pi \alpha_B \epsilon n_H^2} \right)^{1/3}\approx\left( \frac{3 Q_0}{4 \pi \alpha_B \epsilon n_e^2} \right)^{1/3}
\end{equation}
 where $\epsilon$ is the volume filling factor of the gas.  The volume filling factor can be defined by assuming that the gas is structured
 in dense clumps that are surrounded by a lower-density medium.  In this case, the volume filling factor is defined as
\begin{equation}
\epsilon=\frac{\langle n_e \rangle^2}{n_{e,c}^2}
\end{equation}
 where $\langle n_e \rangle$ is the global average electron density and $n_{e,c}$ is the electron density of the clumps.  The volume
 filling factor is equal to unity for a homogeneous constant-density gas, while its value decreases as the density of the clumps
 increases relative to the average density.
  We note that the density estimates from [S~\textsc{ii}] and [O~\textsc{ii}] are based on luminosity-weighted
 measurements of emission line strengths.  Since emission strength scales as the square of the density, we are effectively measuring
 the clump density if a clumpy gas geometry exists, not the global average density \citep{ken84}.  
  Using the definition of the Str\"omgren sphere radius, we can simplify the ionization parameter and resolve its
 dependence on only the rate of ionizing photon production, the electron density, and the volume filling factor.
\begin{equation}\label{eq:uscaling}
\mathcal{U}\propto Q_0^{1/3} n_e^{1/3} \epsilon^{2/3}
\end{equation}
The ionization parameter has a weak dependence on both the rate of ionizing photon production
 and the gas density, and is somewhat more sensitive to the volume filling factor, through which this definition of the
 ionization parameter contains information about the geometry of the gas.

Defining the ionization parameter assuming the geometry of a Str\"omgren sphere is convenient, but likely does not
 hold for real HII regions.  The Str\"omgren geometry assumes a sphere of constant density gas that immediately surrounds the
 central ionizing source.  In local HII regions, feedback from stellar winds can clear out a cavity around the ionizing star cluster
 such that the ionized gas is a shell instead of a filled  sphere \citep[e.g.,][]{wat08}.  Accordingly, the Str\"omgren radius is not necessarily a good
 representative radius for the separation of the illuminated gas and the ionizing source.  It is possible that the wind-blown
 bubble geometry exists at high redshifts where the intensity of star formation is concentrated, or some entirely different geometry
 such as intersecting bubbles.  The scalings presented in
 equation~\ref{eq:uscaling} should then be used with caution because of the breakdown of the Str\"omgren approximation.  Real
 HII regions show a variety of complicated substructure and geometry \citep[e.g.,][]{pel11}.  The Str\"omgren sphere definition of the
 ionization parameter also assumes that the nebula is radiation-bounded (i.e. no hydrogen-ionizing photons escape) instead of
 density-bounded \citep{nak14}, which may not hold true at high redshifts.
  Additionally, in an integrated spectrum the measured ionization parameter is a luminosity-weighted average of all of the
 sources of emission inside the aperture, which includes multiple HII regions and emission from other ISM components.

While the ionization parameter carries interesting information about the ionizing source and gas geometry,
 it can be difficult to determine because it is not
 directly observable, but can only be estimated using calibrations derived from physically motivated models.
The ionization parameter is often estimated using measurements of sets of emission line ratios that
 have some sensitivity to the ionization state of the gas (e.g., lower- and higher-ionization states),
 in conjunction with the predictions of a suite of photoionization models
 \citep{dia00,kew02,dor11,lev14,shi14}.
  Because of differences in the translation between observables and ionization parameter for different
 photoionization models, it is convenient to instead use an empirical emission line ratio as a proxy
 for the ionization parameter.
  Line ratios featuring both higher and lower ionization state transitions from the same element
 can be used to estimate the ionization parameter because of the relation between the ionization parameter
 and the relative populations
 in the two ionization states.  Here, we use
 $O_{32}$=[O~\textsc{iii}]$\lambda\lambda$5007,4959/[O~\textsc{ii}]$\lambda\lambda$3726,3729
 as a proxy for the ionization parameter.
  Systematic uncertainties in this approach result from the way in which a given ionization parameter-sensitive line
 ratio depends also on the shape of the ionization spectrum and the metallicity of the gas, unless these two properties
 can be independently constrained.

\subsection{Sample selection}\label{sec:sample}

In order to study the ionization state of high-redshift galaxies, we selected a sample of
 star-forming galaxies from the MOSDEF parent spectroscopic sample requiring objects to
 fall in the redshift range $2.0<z<2.6$ and have S/N$\geq3$ in [O~\textsc{ii}]$\lambda\lambda$3726,3729,
 H$\beta$, [O~\textsc{iii}]$\lambda$5007, and H$\alpha$.  The flux of [O~\textsc{iii}]$\lambda$4959 is
 taken to be 1/2.98 of the [O~\textsc{iii}]$\lambda$5007 flux \citep{sto00}.  H$\alpha$ and H$\beta$ detections
 were necessary in order to correct line fluxes for dust attenuation, which is important for $O_{32}$ because
 of the large wavelength separation of the emission lines.  Objects with H$\alpha$ and H$\beta$ detections also have robust dust-corrected
 SFRs based on H$\alpha$ luminosities.  AGN were identified and removed based on their
 X-ray and IR properties \citep{coi15} and objects with [N~\textsc{ii}]$\lambda$6584 detected at 3$\sigma$
 or greater were removed if $\log($[N~\textsc{ii}]$\lambda$6584/H$\alpha)>-0.3$.  Any AGN not removed by these selection criteria would
 introduce a bias in emission line ratio diagrams.  We discuss reasons why we are confident that our sample does not contain
 any AGN in Sections~\ref{sec:o32metallicity} and~\ref{sec:bptoffset}.
  Six additional objects were removed because of significant skyline contamination in the relevant emission lines.
  These criteria yield an ionization
 parameter sample of 103 MOSDEF galaxies and the properties of the sample are shown in Table~\ref{tab:o32sample}.
  This sample
 has properties that are nearly identical to those of the parent MOSDEF spectroscopic sample at $z\sim2.3$,
 and is representative of star-forming galaxies with similar stellar masses at this redshift.

\begin{table*}[t]
 \centering
 \caption{Properties of the full ionization parameter sample, and [N~\textsc{ii}] and [S~\textsc{ii}] subsamples}\label{tab:o32sample}
 \begin{tabular*}{0.7\textwidth}{@{\extracolsep{\fill}} c | c c c c c c}
   \hline\hline
   $\ $ & $\langle z\rangle$\tablenotemark{a} & $\sigma_z$\tablenotemark{b} & $\log{(\frac{\mbox{M}_*}{\mbox{M}_\odot})}$\tablenotemark{c} & $\log{(\frac{\mbox{M}_*}{\mbox{M}_\odot})}_{\rm med}$\tablenotemark{d} & SFR\tablenotemark{e} & SFR$_{\rm med}$\tablenotemark{f}\\[6pt] \hline
   Full sample & 2.29 & 0.11 & $8.97-11.22$ & 10.0 & $1.61-228$ & 23.8 \\ 
   {[}N \textsc{ii}] subsample & 2.28 & 0.11 & $9.3-11.22$ & 10.22 & $4.71-228$ & 33.2 \\
   {[}S \textsc{ii}] subsample & 2.29 & 0.11 & $9.3-11.22$ & 10.26 & $5.77-228$ & 33.3 \\
   \hline
 \end{tabular*}
 \tablenotetext{1}{Average redshift of galaxies in the sample.}
 \tablenotetext{2}{Standard deviation of the redshift distribution.}
 \tablenotetext{3}{Range of $\log{(\mbox{M}_*/\msun)}$ of galaxies in the sample.}
 \tablenotetext{4}{Median $\log{(\mbox{M}_*/\msun)}$ of galaxies in the sample.}
 \tablenotetext{5}{Range of SFR in M$_{\odot}$/yr$^{-1}$ of galaxies in the sample, determined from dust-corrected H$\alpha$ luminosity.}
 \tablenotetext{6}{Median SFR in M$_{\odot}$/yr$^{-1}$ of galaxies in the sample, determined from dust-corrected H$\alpha$ luminosity.}
\end{table*}

In the discussion that follows, we will examine multiple line ratio diagrams, some of which involve additional emission lines along with
 oxygen and hydrogen Balmer-series strong lines.  Therefore, we selected subsamples of the ionization parameter
 sample that have additional line detection criteria to plot in these spaces.  We selected a subsample
 of 61 galaxies that additionally have S/N$\geq3$ in [N~\textsc{ii}]$\lambda$6585.
  We also selected a subset of 53 galaxies with S/N$\geq3$ in both
 [N~\textsc{ii}]$\lambda$6585 and [S~\textsc{ii}]$\lambda\lambda$6716,6731.
  The sample properties of the [N~\textsc{ii}] and [S~\textsc{ii}] subsamples are presented in Table~\ref{tab:o32sample}.
  These stringent emission line cuts, requiring all or nearly all of the
 optical strong lines to be detected, introduce a bias against low-mass, low-SFR galaxies.  However, the
 typical galaxy properties of the subsamples are not significantly different from those of the full ionization
 parameter sample or the MOSDEF parent $z\sim2.3$ spectroscopic sample.  While the [N~\textsc{ii}] and
 [S~\textsc{ii}] subsamples have slightly higher \mstar\ and SFR than the full sample, the median values are still
 consistent with these galaxies falling on or near the $z\sim2$ SFR-\mstar\ relation \citep{whi14,shi15}.

We select a sample of typical star-forming galaxies in the local universe from SDSS DR7.
  We require $0.04<z<0.1$ and S/N$\geq$3 in [O~\textsc{ii}]$\lambda\lambda$3726,3729, H$\beta$,
 [O~\textsc{iii}]$\lambda$5007, H$\alpha$, and [N~\textsc{ii}]$\lambda$6585.  Once again, the flux of
 [O~\textsc{iii}]$\lambda$4959 is assumed to be equal to 1/2.98 of the [O~\textsc{iii}]$\lambda$5007
 flux \citep{sto00}.  A detection in [N~\textsc{ii}] is required because AGN are rejected using
 the demarcation of \citet{kau03b} in the [O~\textsc{iii}]/H$\beta$ vs. [N~\textsc{ii}]/H$\alpha$ diagram.
  These criteria yield a local sample of 68,453 star-forming galaxies at $z\sim0.07$.  We also selected a subsample
 of 65,000 local galaxies that additionally have S/N$\geq3$ in [S~\textsc{ii}]$\lambda\lambda$6716,6731.

\subsection{O$_{32}$ and global galaxy properties}

\begin{figure}
 \centering
 \includegraphics[width=\columnwidth]{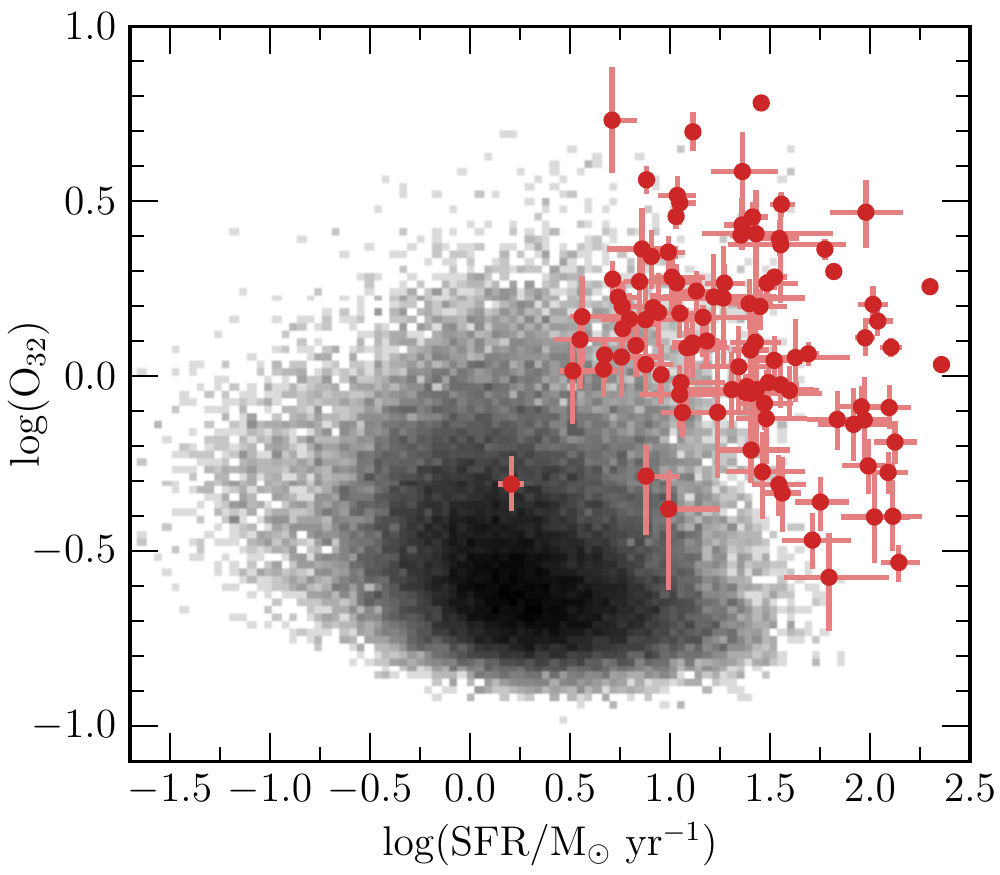}
 \includegraphics[width=\columnwidth]{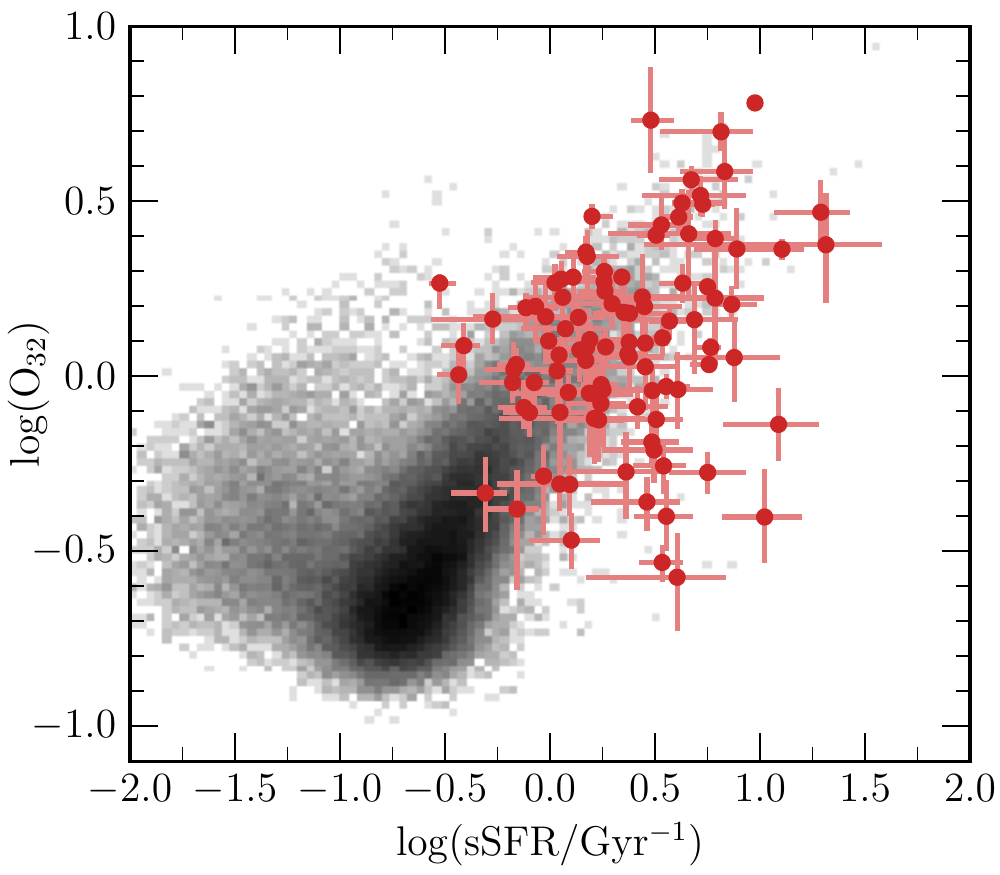}
 \caption{O$_{32}$ vs. SFR (top) and sSFR (bottom) for local star-forming galaxies from SDSS (gray histogram)
 and $z\sim2.3$ star-forming galaxies from MOSDEF (red circles).
}\label{fig:o32sfr}
\end{figure}

\begin{figure*}
 \centering
 \includegraphics[width=0.8\textwidth]{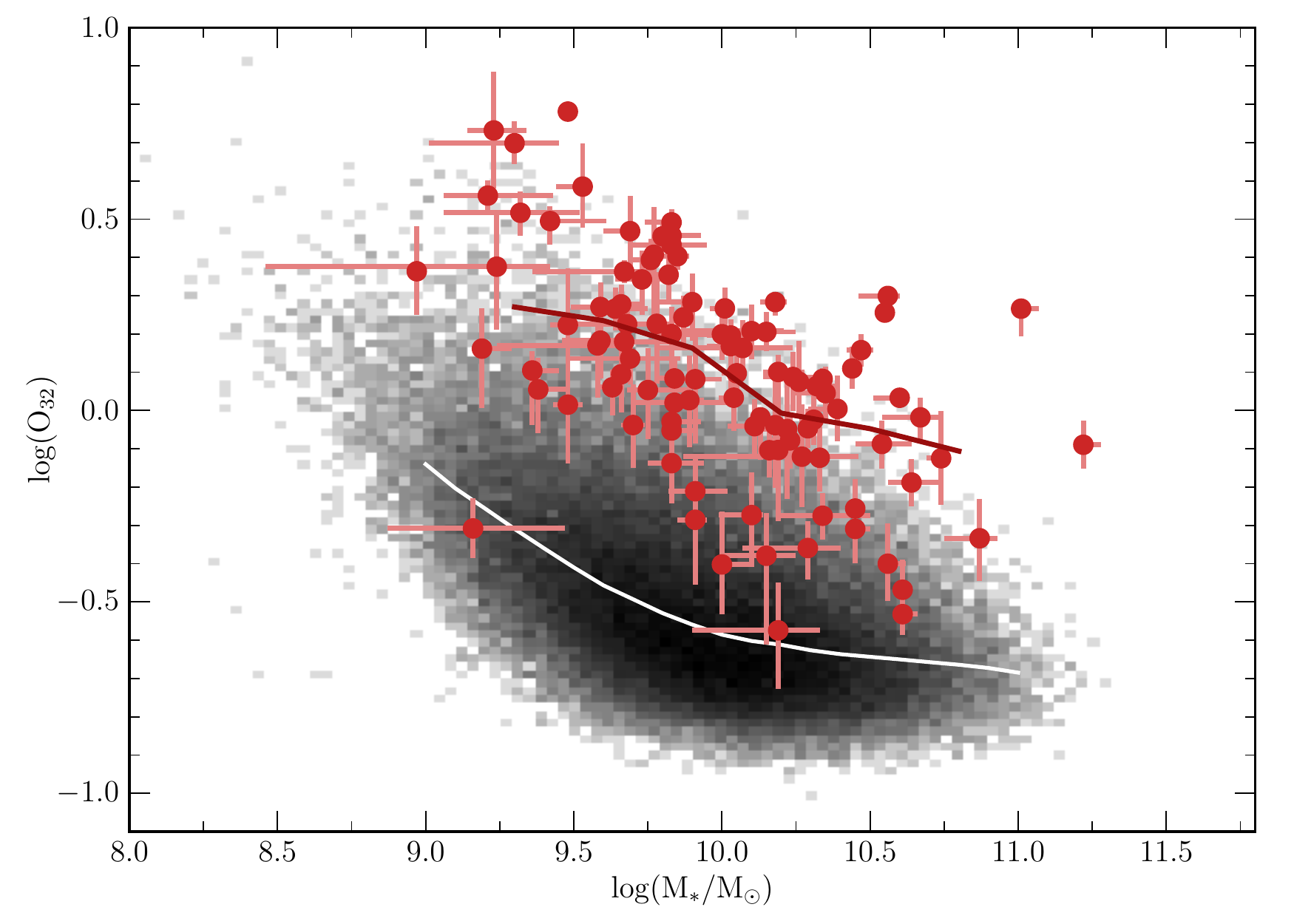}
 \caption{O$_{32}$ vs. stellar mass for local star-forming galaxies from SDSS (gray histogram)
 and $z\sim2.3$ star-forming galaxies from MOSDEF (red circles).  The solid white line shows the
 running median O$_{32}$ in bins of stellar mass for local galaxies.  The solid dark red line
 shows the running median of $z\sim2.3$ galaxies.
}\label{fig:o32mass}
\end{figure*}

Many studies have suggested that $z\gtrsim2$ galaxies have systematically higher ionization parameters
 than are typical for local galaxies \citep{hol14,nak13,nak14,shi14,ste14}.  However, it is imperative to consider
 the evolution in global galaxy properties with redshift when interpreting the apparently high
 ionization parameters observed at high redshifts.  To this end, we investigate the dependence of the
 ionization parameter on global galaxy properties through the proxy of O$_{32}$, motivated by the
 comparisons performed in \citet{nak14}.  Figure~\ref{fig:o32sfr} presents O$_{32}$ vs. SFR and sSFR
 for local star-forming galaxies (gray histogram) and the $z\sim2.3$ ionization parameter
 sample (red circles), while Figure~\ref{fig:o32mass} shows the dependence of O$_{32}$ on
 stellar mass for the same samples.  As described in Section~\ref{sec:densgalprop}, the $z\sim2.3$ galaxies span a similar range in \mstar\ as
 local SDSS galaxies, but have significantly higher SFR and sSFR at a given stellar mass, consistent
 with the evolution of the SFR-\mstar\ relation with redshift \citep{red12,whi14,shi15}.

Figure~\ref{fig:o32sfr}, top panel, shows that there is a weak anti-correlation between O$_{32}$
 and SFR for local galaxies, although there is significant scatter in O$_{32}$ at fixed SFR.
  Performing a Spearman correlation test yields a correlation coefficient of -0.19, indicating a weak anti-correlation,
 with a p-value\footnote[8]{In the
 Spearman correlation test, the p-value represents the probability of the dataset being drawn from an uncorrelated underlying
 distribution.  A correlation or anti-correlation with a p-value less than 0.003 has a significange greater than 3$\sigma$.}
 that is essentially zero due to the large sample size from SDSS.
  It is known that
 there is an anti-correlation between metallicity and ionization parameter in the local universe
 \citep{dop86,dop06a,dop06b,per14,sanchez15},
 while SFR and metallicity are correlated with large scatter \citep{man10,lar13}.
  Therefore, SFR and ionization parameter are anti-correlated (with large scatter) and the top
 panel of Figure~\ref{fig:o32sfr} confirms that O$_{32}$ acts as a proxy for the ionization parameter.
  The $z\sim2.3$ galaxies also show a weak anti-correlation between O$_{32}$ and SFR, with a Spearman correlation
 coefficient of -0.30 and a p-value of 0.002.

In the bottom panel of Figure~\ref{fig:o32sfr}, there is a tight correlation between sSFR and O$_{32}$ for
 $z\sim0$ galaxies, with a Spearman correlation coefficient of 0.42.
  The bulk of local star-forming galaxies lie on this relation, while a small fraction of local  galaxies
 with the least star formation ($\log(\mbox{sSFR})\lesssim-1.0$) do not show any correlation.  It is possible that
 these low-sSFR galaxies have a very low level of ongoing star formation and are
 transitioning to the red sequence via secular evolution or some
 other mechanism associated with the cessation of star formation.  The ISM conditions in these low-sSFR galaxies may
 be different from those in typical local star-forming galaxies and would not be expected to follow the
 same trends.  While $z\sim2.3$ galaxies inhabit much of the same parameter space as highly star-forming
 local galaxies, they do not exhibit the same tight correlation between sSFR and O$_{32}$.  Performing a Spearman
 correlation test yields a correlation coefficient of 0.28 with a p-value of 0.004, indicating a weak but
 significant correlation.  There is significant
 scatter in O$_{32}$ at fixed sSFR, and a non-negligible fraction of the high-redshift sample has high sSFR
 and low O$_{32}$, a region of the parameter space where essentially no local galaxies are found.  The difference
 between $z\sim0$ and $z\sim2.3$ galaxies in this space is interesting, but outside of the scope of this investigation.

The most intriguing of these diagrams is that of O$_{32}$ vs. \mstar, shown in Figure~\ref{fig:o32mass}.  Local galaxies
 show a clear anti-correlation between O$_{32}$ and \mstar\ that is fairly tight, with a Spearman correlation
 coefficient of -0.52.  In order to make the local
 trend more clear, Figure~\ref{fig:o32mass} shows the running median O$_{32}$ at a given \mstar\ as a white line.
  This anti-correlation is consistent with the existence of a tight correlation between \mstar\ and
 metallicity, known as the mass-metallicity relation \citep[MZR;][]{leq79,tre04,kew08,man10,and13}, and an
 anti-correlation between metallicity and ionization parameter \citep{per14,sanchez15}.  In fact, O$_{32}$
 has been shown to be a metallicity indicator for objects up to $z\sim0.8$ \citep{mai08,jon15}.
  The relation between O$_{32}$ and \mstar\ appears to flatten out at high stellar
 masses, consistent with the observed behavior of the local MZR.

The $z\sim2.3$ sample also displays a fairly tight
 relation between O$_{32}$ and \mstar such that higher \mstar\ corresponds to lower O$_{32}$, with a Spearman
 correlation coefficient of -0.57 and a p-value of $3.7\times10^{-10}$.  We show the running
 median O$_{32}$ of the high-redshift galaxies at a given \mstar\ as a dark red line in Figure~\ref{fig:o32mass}.
  The high-redshift anti-correlation shows nearly the same slope as that of the local relation, only offset towards
 higher O$_{32}$ at fixed \mstar.  We find that $z\sim2.3$ galaxies have O$_{32}$ values $\sim0.6$ dex higher
 at a given \mstar, suggesting that high-redshift galaxies have significantly higher ionization parameters than
 local galaxies of the same stellar mass if the translation between O$_{32}$ and ionization parameter is the same
 at both redshifts.  Stellar mass and metallicity are correlated at $z\sim2.3$ as well \citep[e.g.,][]{erb06,mai08,ste14,san15}, but the MZR evolves
 such that galaxies at a given \mstar\ have lower metallicities than are observed locally.  The existence
 of a clear O$_{32}$ vs. \mstar\ anti-correlation  at $z\sim2.3$ is suggestive of an anti-correlation between metallicity
 and ionization parameter existing at high redshifts as well.

The striking similarity of the shape of the O$_{32}$ vs. \mstar\ relation for local and $z\sim2.3$ galaxies
 suggests that a similar mechanism may set the observed ionization parameter at both redshifts, but must
 evolve with redshift such that high-redshift galaxies have higher ionization parameters at a given stellar mass.
  The evolution of the MZR from $z\sim0$ to $z\sim2.3$ provides a natural explanation for the apparent change
 in ionization parameter at fixed stellar mass.
  We further investigate the interplay of O$_{32}$, \mstar, and metallicity by employing the use of emission
 line ratio diagrams.

\subsection{O$_{32}$ and metallicity}\label{sec:o32metallicity}

We have shown that there is a relation between O$_{32}$ and \mstar, the shape of which is similar for local and
 $z\sim2.3$ galaxies, that has a higher normalization in O$_{32}$ at high redshift.  However, stellar mass is a global
 property that is not directly related to the production of emission lines in individual star-forming regions
 that are an observable probe of the ionization parameter.  The metallicity of the gas in star-forming regions,
 on the other hand, has a direct impact on both the ionizing spectrum, assuming the gas-phase metallicity is
 related to the stellar metallicity of the ionizing cluster, and the intrinsic emission line fluxes.  Therefore,
 comparing O$_{32}$ values at fixed metallicity rather than fixed \mstar\ utilizes a property that directly influences the
 physical conditions in star-forming regions, including ionization parameter, and removes systematic effects
 introduced by the evolution of galaxy scaling relations with \mstar.

\begin{figure*}
 \centering
 \includegraphics[width=0.495\textwidth]{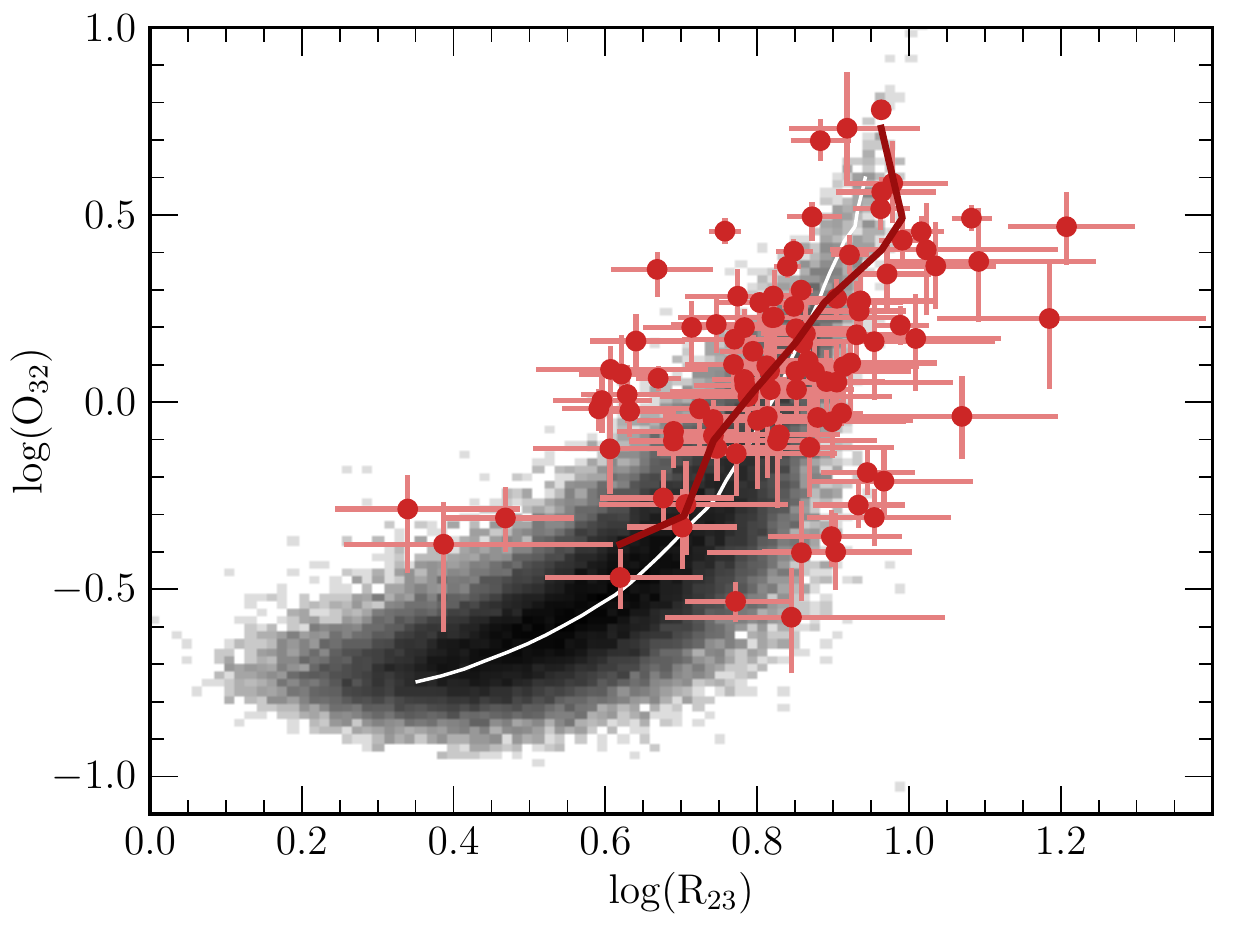}
 \includegraphics[width=0.495\textwidth]{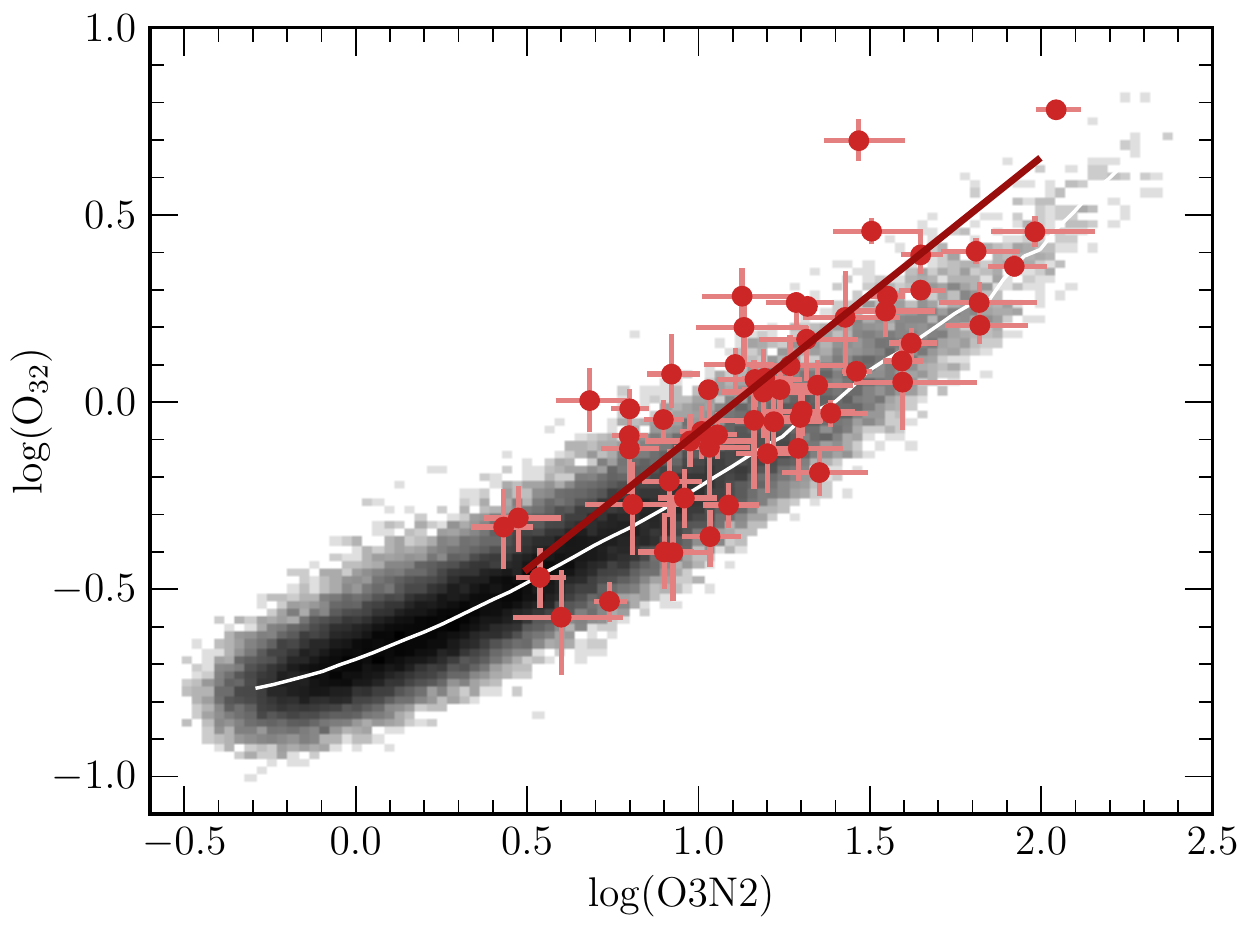}
 \caption{O$_{32}$ vs. R$_{23}$ (left panel) and O3N2 (right panel).  The gray histogram shows the distribution
 of local star-forming galaxies from SDSS.  The red points and error bars denote $z\sim2.3$ star-forming galaxies from the MOSDEF
 survey.  In both panels, the white line shows the running median of the local sample.  The dark red line shows the running median
 of the $z\sim2.3$ in the left panel, and the best-fit linear relation to the $z\sim2.3$ galaxies in the right panel.
  The right panel only includes galaxies in the [N~\textsc{ii}] subsample.
}\label{fig:o32vsr23o3n2}
\end{figure*}

Figure~\ref{fig:o32vsr23o3n2} shows the dependence of O$_{32}$ on R$_{23}$ (left) and O3N2 (right) for local
 SDSS galaxies (gray histogram) and the $z\sim2.3$ ionization parameter sample.  Note that the right-hand
 panel only includes high-redshift galaxies in the [N~\textsc{ii}] subsample since the O3N2 ratio requires
 [N~\textsc{ii}]$\lambda$6584.
  Figure~\ref{fig:o32vsr23o3n2} demonstrates that the $z\sim2.3$ sample is likely free from
 AGN contamination.  AGN are found at low O3N2, high R$_{23}$, and
 high O$_{32}$.  No galaxies in the $z\sim2.3$ sample fall on the AGN sequence in the O3N2 diagram.  In the R$_{23}$ 
 diagram, there are several $z\sim2.3$ galaxies in the same region of parameter space as the AGN sequence ($\log($R$_{23})>0.95$),
 but these galaxies fall below the \citet{kau03b} line in the [O~\textsc{iii}]/H$\beta$ vs. [N~\textsc{ii}]/H$\alpha$ diagram
 and tend to have large uncertainties in R$_{23}$.  

  The R$_{23}$ index is sensitive to metallicity, but is double-valued \citep{kew02}.  This problem can be overcome by using a second
 excitation-sensitive line ratio in
 tandem with R$_{23}$.  \citet{sha15} used direct-method abundances of stacks of local galaxies \citep{and13} to show that
 the local sequence in the O$_{32}$ vs. R$_{23}$ diagram is a sequence in monotonically increasing
 metallicity from the high-O$_{32}$, high-R$_{23}$, high-excitation tail towards the low-O$_{32}$,
 low-R$_{23}$, low-excitation region.
  The O3N2 index is also a metallicity indicator, reflecting the fact that
 the narrow sequence of star-forming galaxies in the BPT diagram is also a sequence in metallicity,
 and has been used as such in empirical metallicity calibrations \citep[e.g.,][]{pet04}.
  The position on either of these diagrams reflects the ionization
 parameter at a given oxygen abundance.

Using a smaller sample from the early MOSDEF dataset, \citet{sha15} showed that $z\sim2.3$ star-forming
 galaxies seem to follow the same distribution as local galaxies in the low-metallicity, high-excitation
 tail of the O$_{32}$ vs. R$_{23}$ diagram.  We confirm this finding with the current MOSDEF $z\sim2.3$ sample.
  The $z\sim2.3$ galaxies display no systematic offset with respect to the local galaxies, and the bulk
 of the high-redshift sample inhabits the region in which $12+\log(\mbox{O/H})\lesssim8.6$ for local galaxies.
  The running median of the $z\sim2.3$ sample (dark red line) closely follows that of $z\sim0$ galaxies (white line)
 in the O$_{32}$ vs. R$_{23}$ diagram.\footnote[9]{
A running median creates a smooth representation of the local sequence of star-forming
 galaxies because of the large number of galaxies in the local sample.  The small number of galaxies in the $z\sim2.3$
 sample and subsamples can cause the running median to be uneven and erratic.
  A linear fit better represents the $z\sim2.3$ subsample distributions in the
 O3N2 diagram, while a running median is still used in the R$_{23}$ diagram because the O$_{32}$ vs. R$_{23}$ sequence
 displays curvature.}
  We also find that $z\sim2.3$ galaxies closely follow the subsolar abundance tale of the local distribution
 in the O$_{32}$ vs. O3N2 diagram, although there is a slight systematic offset towards lower O3N2 at fixed O$_{32}$.
  By comparing the best-fit linear relation of the $z\sim2.3$ sample (dark red line) to the running median of the $z\sim0$ sample
 (white line), we find the mean value of this offset to be 0.23 dex in O3N2 at fixed O$_{32}$.
  We will discuss the offset in the O3N2 diagram further in Section~\ref{sec:zo32}.

We propose the following scenario to explain the observed position of high-redshift galaxies in
 the O$_{32}$ vs. R$_{23}$, O$_{32}$ vs. O3N2, and O$_{32}$ vs. \mstar\ diagrams.
  Taken together, the two plots in Figure~\ref{fig:o32vsr23o3n2} suggest
 that the ionization parameter
 at fixed metallicity is the same for galaxies at $z\sim2.3$ and $z\sim0$.  Galaxies at high redshift
 must then follow the same anti-correlation between oxygen abundance and ionization parameter that is observed
 locally.  This $\mathcal{U}$ vs. O/H relation shows little to no redshift evolution between $z\sim0$ and $z\sim2.3$.
  The nearly constant offset towards higher O$_{32}$ at fixed \mstar\ observed in Figure~\ref{fig:o32mass} is
 then simply a consequence of the evolution of the MZR with redshift.  In \citet{san15}, we found
 $z\sim2.3$ galaxies have metallicities $\sim0.3$ dex lower than local galaxies at fixed \mstar\ using the
 O3N2 calibration of \citet{pet04}.  If $z\sim0$ and $z\sim2.3$ galaxies follow the same metallicity-ionization
 parameter relation, then the decrease in metallicity at fixed \mstar\ leads to an increase in O$_{32}$, as observed
 in Figure~\ref{fig:o32mass}.

Earlier results suggesting that high-redshift galaxies have higher ionization
 parameters than are seen locally were based on either a comparison at fixed stellar mass \citep[e.g.,][]{hol14}
 or a comparison of the average ionization parameter of the entire local star-forming population to that of
 high-redshift galaxies \citep[e.g.,][]{hai09}.  At fixed metallicity, which is more directly related to the
 ionization state of the gas, we find that $z\sim2.3$ galaxies have roughly the same ionization parameters
 compared to local galaxies.  The ionization state of star-forming regions in galaxies in our high-redshift sample must
 then be similar to what is observed in metal-poor local galaxies.

In Section~\ref{sec:dens}, we found that the density of star-forming regions increases significantly from $z\sim0$ to $z\sim2.3$.
  Here, we are suggesting that galaxies at $z\sim0$ and $z\sim2.3$ have the same ionization parameter at fixed metallicity
 despite the difference in density.  This initially seems to be at odds with the scaling of ionization parameter with
 density presented in equation~\ref{eq:uscaling}.  An increase in density of an order of magnitude would correspond to an increase
 in the ionization parameter of more than a factor of two.  \citet{shi14} used the relation between ionization parameter
 and electron density to explain the high ionization parameters observed at $z\sim2-3$ by an increase in the density of
 star-forming regions.  If the scalings
 in equation~\ref{eq:uscaling} hold at $z\sim0$ and $z\sim2.3$, then an increase in density does not guarantee an increase
 in the ionization parameter unless the volume filling factor is the same at both redshifts.  If high-redshift HII regions
 are clumpier, the volume filling factor would decrease with redshift and could offset the effect of an increase in density.
  However, the scalings presented in equation~\ref{eq:uscaling}
 are derived assuming the Str\"omgren approximation, which does not apply to many local HII regions and likely does not
 hold at $z\sim2.3$.  Thus, an increase of a factor of 10 in density from $z\sim0$ to $z\sim2.3$ does not necessitate
 higher ionization parameters at fixed metallicity.
 
Our proposed scenario is only valid if some assumptions hold true.  A location in O$_{32}$ vs. R$_{23}$ space must
 correspond to the same metallicity regardless of redshift.  The relation between O3N2 and metallicity must
 not evolve significantly with redshift.  Finally, the translation between ionization parameter and O$_{32}$
 must not evolve significantly from $z\sim0$ to $z\sim2.3$.  We investigate the validity of these assumptions in the
 following section.

\section{Discussion}\label{sec:discussion}

In this section, we present evidence in support of our proposed scenario that $z\sim2.3$ star-forming
 galaxies have roughly the same ionization parameter as $z\sim0$ galaxies when comparing at fixed metallicity.  In
 Section~\ref{sec:qo32}, we investigate whether the translation between O$_{32}$ and ionization
 parameter evolves with redshift using a set of simple photoionization models in combination with the observed
 position of $z\sim2.3$ galaxies in emission line ratio diagnostic diagrams.
  We explore whether or not there is evidence that the metallicity dependence of R$_{23}$ and O3N2 changes
 significantly from $z\sim0$ to $z\sim2.3$ in Section~\ref{sec:zo32}.  In Section~\ref{sec:qohanticorr},
 we discuss additional evidence for the existence of an anti-correlation between O/H and $\mathcal{U}$ and how the
 existence of this anti-correlation affects the interpretation of photoionization model grids.
  In Section~\ref{sec:bptoffset}, we use our results to explore the cause of the well-known offset of $z\sim2$ galaxies
 in the [N~\textsc{ii}] BPT diagram compared to local galaxies.
  Finally, we discuss the uncertainty that diffuse ionized gas introduces in the interpretation of emission line ratios
 from integrated-light galaxy spectra in Section~\ref{sec:diffuse}.

\subsection{Is the ionization parameter-O$_{32}$ relation redshift invariant?}\label{sec:qo32}

We first address whether or not the relationship between ionization parameter and
 O$_{32}$ is the same at $z\sim0$ and $z\sim2.3$.
  We describe the simple photoionization models used for this analysis, discuss the
 interdependence of the shape of the ionizing spectrum, ionization parameter, and O$_{32}$,
 and estimate the typical ionization parameter of $z\sim2.3$ and local galaxies based on
 the models.

\subsubsection{Description of the Cloudy photoionization models}

As previously mentioned, the chief
 difficulty in determing the ionization parameter is that it can only be done with reference to a specific
 set of photoionization models.  The extent to which the value of the estimated ionization parameter
 can be trusted depends on how well the models represent the observed objects and
 can produce self-consistent predictions in multiple line ratio spaces.  Given the
 uncertainty of photoionization models both locally and at high-redshift where the
 physical properties are less constrained, we use a suite of simple photoionization
 models to understand qualitatively the trends in emission line ratios when changing
 the different input parameters of the models.  We do not, however, use these models
 to place tight constraints on the metallicity or ionization parameter of any local or high-redshift
 galaxies.

We use the photoionization code
 Cloudy\footnote[10]{Calculations were performed with version 13.02 of Cloudy, last described by \citet{fer13}.}
 to model emission line ratios from star-forming regions with a range of physical conditions.
  These models are very similar to those used by \citet{ste14} to investigate the position of
 $z\sim2.3$ galaxies in the [O~\textsc{iii}]/H$\beta$ vs. [N~\textsc{ii}]/H$\alpha$ diagram.
  There are five main input parameters that determine the location of a grid point in various
 emission line ratio diagrams: hydrogen gas density, gas-phase metallicity, ionization parameter,
 shape of the ionizing spectrum, and N/O abundance ratio.

In Section~\ref{sec:dens}, we presented
 a robust characterization of the typical electron density in local and $z\sim2.3$ galaxies, finding densities
 of $\sim25$ cm$^{-3}$ and $\sim250$ cm$^{-3}$, respectively.  In HII regions, the gas is fully ionized and
 the electron density provides a good estimate of the hydrogen gas density.
  Since we have measured the typical
 density at $z\sim0$ and $z\sim2.3$, we only allow the density to be either 25 cm$^{-3}$ or 250 cm$^{-3}$ in the models.
  The metallicity sets the global abundance scale of the gas, which is assumed to follow a solar abundance
 pattern with the exception of nitrogen.  We vary the metallicity between 0.2 Z$_{\odot}$ and 1.0 Z$_{\odot}$
 ($12+\log(\mbox{O/H})=8.0-8.69$)
 in 0.2 Z$_{\odot}$ steps with solar metallicity corresponding to $12+\log(\mbox{O/H})_{\odot}=8.69$ \citep{asp09}.
  The ionization parameter sets the ionization state of the gas and is allowed to vary between
 $\log(\mathcal{U})=-3.6$ and $-1.5$ ($\log(\frac{q}{\rm cm/s})=6.9-9.0$) in 0.1 dex steps.

As described in Section~\ref{sec:ionpdef}, the shape of the ionizing
 spectrum affects the relative populations of an element in different ionized states.  A harder ionizing spectrum
 results in a larger fraction of oxygen in O~\textsc{iii} compared to O~\textsc{ii}, for example.  We use
 a blackbody spectrum with an effective temperature of 40,000 K, 50,000 K, or 60,000 K as the input ionizing spectrum.
  \citet{ste14} showed that, when properly normalized, a blackbody spectrum is a good approximation of
 the spectrum of massive stars bluewards of 912 \AA\ using BPASS stellar models that include effects
 from binarity \citep{bpass1,bpass2}.  We have found this observation to hold true when using
 Starburst99 (SB99) stellar models that include effects of rotation in massive stars \citep{sb99}.  We note that
 the effective temperature of the blackbody is not the same as the effective temperature of a star.
  It is simply a parameter that allows us to specify the shape of the input spectrum.  When referring to a
 ``harder" ionizing spectrum, we are referring to an increase in the blackbody effective temperature of the
 input spectrum.  We additionally utilize input spectra produced by SB99 using the Geneva 2012/13 tracks \citep{eks12,geo13}
 that include the effects of rotation in massive stars \citep{sb99}.  We create two input spectra from SB99
 that create bracketing cases of a very hard ionizing spectrum, which we refer to as ``\textit{SB99 hard}", and a softer
 ionizing spectrum, which we refer to as ``\textit{SB99 soft}."  \textit{SB99 hard} is produced assuming a single burst of star-formation
 that formed 0.5~Myr ago with stellar metallicity of 1/7~Z$_{\odot}$.  \textit{SB99 soft} instead
 assumes a 10~Myr-old population with solar metallicity formed with a continuous SFR of 1~M$_{\odot}$/yr.  An age of
 10~Myr was chosen to ensure that the ionizing spectrum of the stellar population had reached a steady state, occuring after
 $\sim5$~Myr \citep{kew01}.  In both
 cases, a \citet{sal55} IMF slope is assumed above 0.5~M$_{\odot}$.  These two cases roughly bracket the
 range of ionizing spectra appropriate for HII regions contributing significantly to integrated-light galaxy spectra.

In reality, the shape of the ionizing spectrum should be related to the gas-phase metallicity,
 which traces the stellar metallicity since recently formed massive stars are ionizing the remnants of their
 birth cloud.  In lower metallicity stars, there is less metal line blanketing and opacity in the
 stellar atmospheres, leading to hotter effective temperatures and harder ionizing spectra.  However, we
 allow the shape of the ionizing spectrum to vary separately from the metallicity to accomodate
 the possibility that the hardness of the ionizing spectrum at fixed metallicity evolves with redshift.

A solar abundance pattern is assumed for all elements except nitrogen.  In the local universe, the N/O
 abundance ratio is observed to have a dependence on O/H, such that N/O is a constant value at low
 abundance but begins to rise roughly linearly with O/H at higher abundance \citep{per09,pil12,and13,per14}.  At low metallicity, nitrogen
 is a primary nucleosynthetic product of hydrogen and helium burning.  At high
 metallicity, nitrogen is produced as a secondary product through the CNO cycle where the yield of nitogren
 depends on the amount of pre-existing C and O, which leads to the
 dependence of N/O on O/H \citep{van06}.  There is disagreement about the shape of the N/O vs. O/H relation
 \citep[see Fig. 12 in ][]{ste14}.  We assume the relation found by \citet{per09}, which is a simple linear
 relation over the range of metallicities considered in the models.  We note that the assumed N/O ratio
 can strongly affect those line ratios involving [N~\textsc{ii}]$\lambda$6584, but has
 negligible effects on other line ratios.

\subsubsection{Ionization parameter and the hardness of the ionizing spectrum at $z\sim0$ to $z\sim2.3$}

\begin{figure*}
 \centering
 \includegraphics[width=\textwidth]{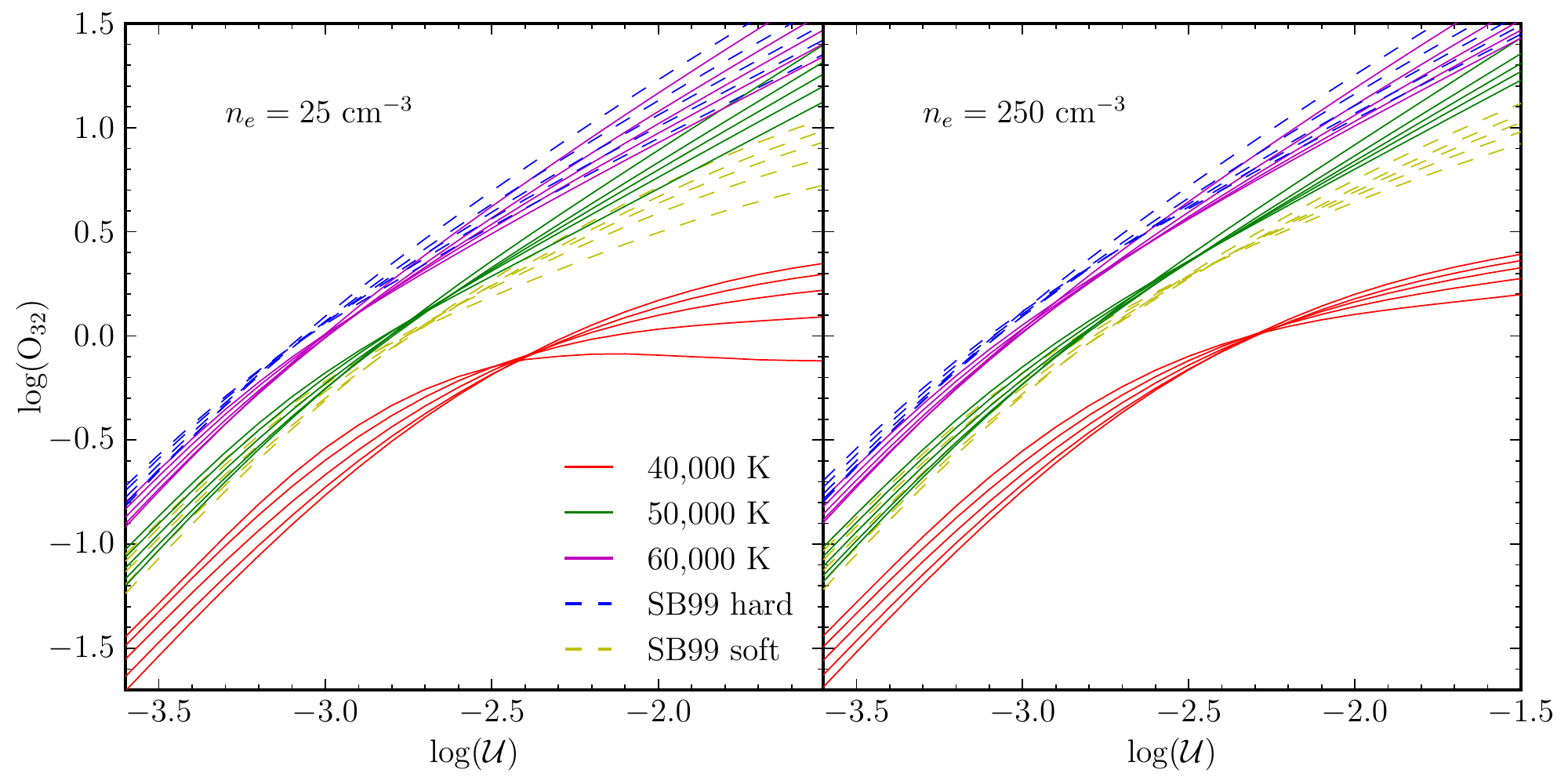}
 \caption{O$_{32}$ vs. ionization parameter, $\mathcal{U}$, from simple photoionization models of HII regions assuming
 a gas density of 25~cm$^{-3}$ (top panel) and 250~cm$^{-3}$ (bottom panel).  The dependence of O$_{32}$ on $\mathcal{U}$
 is extracted from a set of models assuming different ionizing spectra, denoted by the color of the line, and different
 gas-phase metallicities.  Input ionizing spectra are assumed to be either a blackbody spectrum with an effective
 temperature of 40,000-60,000~K or one of two spectra produced by Starburst99 (\textit{SB99 hard} and \textit{SB99 soft}).  Each
 line of a single color connects models with the same gas-phase metallicity, from 0.2~Z$_{\odot}$ to 1.0~Z$_{\odot}$.
}\label{fig:qo32}
\end{figure*}

From the set of models with density, gas-phase metallicity, ionization parameter, ionizing spectrum, and N/O ratio
 defined as above, we extract the relationship between O$_{32}$ and ionization parameter to
 resolve its dependence on the various input parameters.  Figure~\ref{fig:qo32} shows O$_{32}$ vs.
 $\log(\mathcal{U})$ for models with electron density of 25~cm$^{-3}$ (top panel) and 250~cm$^{-3}$
 (bottom panel).  The line color denotes the input spectrum, while the different lines of a single color
 connect grid points with the same metallicity.  We note that the \textit{SB99 hard} models behave very similarly
 to the 60,000~K blackbody models.  The \textit{SB99 soft} spectrum appears to be slightly softer than
 a 50,000~K blackbody based on the position of the model grids.
  Variation in N/O has no effect on the relationship between
 O$_{32}$ and $\mathcal{U}$.  In addition, Figure~\ref{fig:qo32} shows that this relation has
 very little dependence on the gas density.  Lines of
 constant metallicity show that the relation between O$_{32}$ and $\mathcal{U}$ has only a small
 dependence on gas-phase metallicity \textit{when the shape of the
 input spectrum is fixed}.  The one exception is for a soft ionizing spectrum with high ionization parameter
 and low gas density (T$_{\rm eff}=40,000$~K, $\log(\mathcal{U})>-2.5$, $n_e=25$~cm$^{-3}$),
 a region in which real objects are unlikely to be found.
  Calibrations of O$_{32}$ and the ionization parameter typically show significant
 dependence on metallicity because the ionizing spectrum is tied to the metallicity \citep[e.g.,][]{kew02}.
  On the other hand, the hardness of the ionizing spectrum has a significant effect on the O$_{32}$ vs.
 $\log(\mathcal{U})$ relation, such that a harder ionizing spectrum produces a larger O$_{32}$ value at
 fixed ionization parameter.  Therefore, the translation between O$_{32}$ and ionization parameter will
 only show significant evolution if the hardness of the ionizing spectrum at a given metallicity evolves
 with redshift.

We investigate the possibility of the shape of the ionizing spectrum evolving with redshift by combining
 the observed line ratios for our $z\sim2.3$ sample
 with grids from the Cloudy models in emission line ratio diagrams.
  It has been proposed that the position of high-redshift galaxies in the [O~\textsc{iii}]/H$\beta$ vs.
 [N~\textsc{ii}]/H$\alpha$ diagram can be explained by a systematically harder ionizing spectrum
 compared to that of local galaxies with similar metallicity \citep{kew13t, ste14}.  If true, this
 explanation would lead to evolution of the ionization parameter-O$_{32}$ relation.

\begin{figure*}
 \centering
 \includegraphics[width=\textwidth]{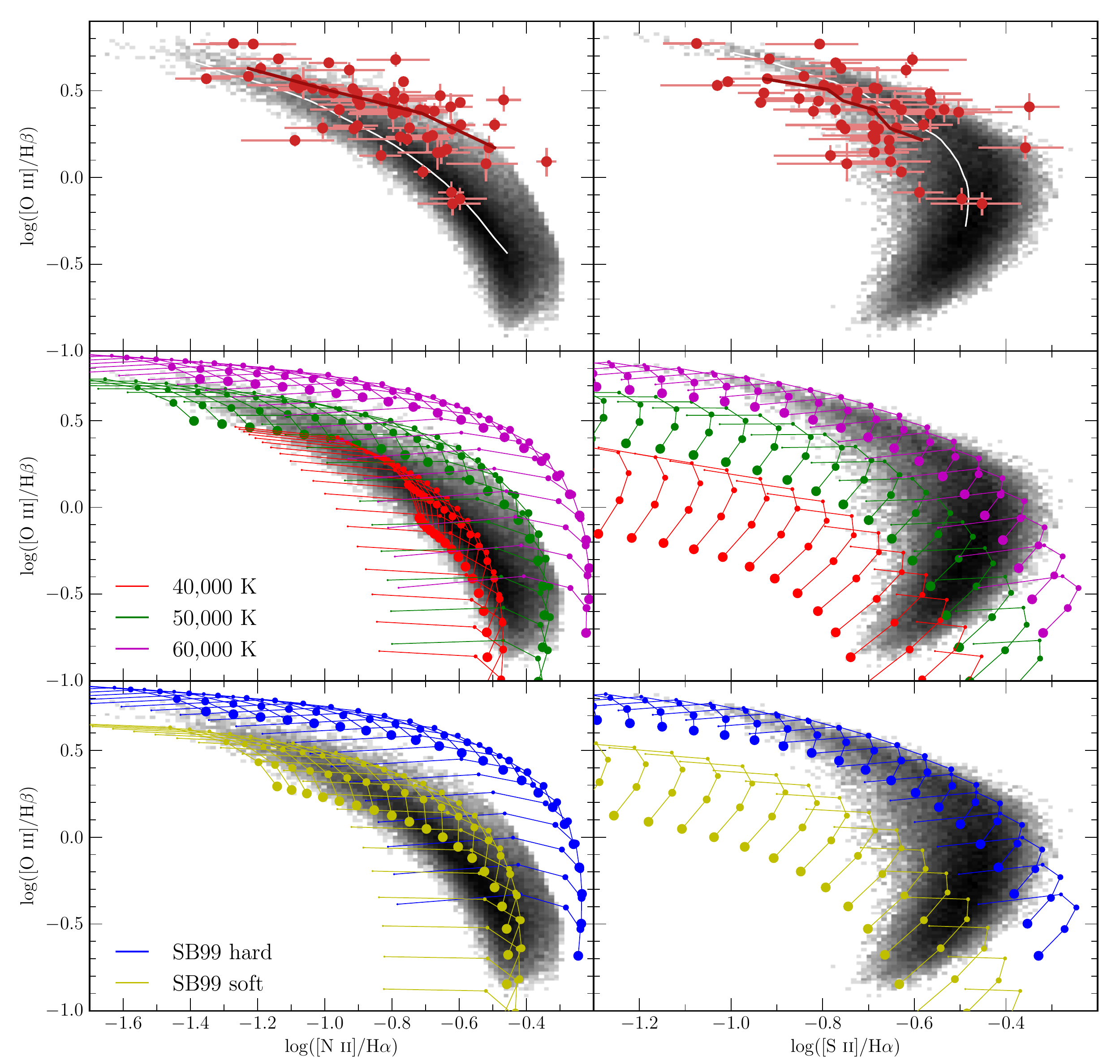}
 \caption{[N~\textsc{ii}] (left column) and [S~\textsc{ii}] (right column) BPT diagrams.  The local sequence of star-forming galaxies from
 SDSS is shown as the gray histogram in all panels.  The top row shows the position of $z\sim2.3$ star-forming galaxies in
 the [N~\textsc{ii}] and [S~\textsc{ii}] BPT diagrams as red points with error bars.  The white and dark red lines show the
 running median of the local and $z\sim2.3$ samples, respectively.
  The bottom two rows show grids from simple photoionization
 models of HII regions produced using Cloudy.  Different colors represent different input ionizing spectra.  In the middle row,
 the input ionizing spectrum is assumed to be a blackbody spectrum and the effective temperature is listed, representing the
 hardness of the ionizing spectrum.  In the bottom row, the input ionizing spectrum is assumed to be one of two spectra produced
 by the stellar population synthesis code Starburst99 (\textit{SB99 hard} and \textit{SB99 soft}).  Colored circles show the model
 grid points, where the size of the circle represents the gas-phase metallicity, from 0.2~Z$_{\odot}$(smallest circles) to solar
 metallicity (largest circles).  Grid points of constant ionization parameter with the same input ionizing spectrum are connected
 with solid lines.  All model grids shown are calculated assuming a hydrogen gas density of 250 cm$^{-3}$.
}\label{fig:bptmodels}
\end{figure*}

Figure~\ref{fig:bptmodels} shows the [O~\textsc{iii}]/H$\beta$ vs. [N~\textsc{ii}]/H$\alpha$ ([N~\textsc{ii}]
 BPT) diagram (left column) and the [O~\textsc{iii}]/H$\beta$ vs. [S~\textsc{ii}]/H$\alpha$ ([S~\textsc{ii}]
 BPT) diagram (right column).  The local distribution of star-forming galaxies is shown as the gray histogram
 in all panels.  The top row presents the observed line ratios for the $z\sim2.3$ [N~\textsc{ii}] subsample
 (top left) and [S~\textsc{ii}] subsample (top right).  The middle and bottom rows show the Cloudy model grid points as
 circles.  The size of the circle indicates the metallicity, with the largest size indicating solar metallicity
 and the smallest size indicating 0.2 Z$_{\odot}$.  The color indicates the input ionizing spectrum, with the
 effective temperature listed for blackbodies in the middle row and the bracketing SB99 models shown in the bottom row.
  Solid lines connect points of constant ionization parameter,
 with ionization parameter increasing to the upper left.  The model grids are only shown for a density of
 250 cm$^{-3}$.  Displaying models at a single density will suffice since we are only interested in discerning
 trends in line ratio with the model input parameters instead of making quantitative predictions.  While the
 absolute line ratios change, the trends with metallicity, ionization parameter, and hardness of the ionizing
 spectrum are the same regardless of the assumed density.  Qualitatively, increasing the electron density while keeping
 all other parameters fixed increases [O~\textsc{iii}]/H$\beta$, [N~\textsc{ii}]/H$\alpha$, and [S~\textsc{ii}]/H$\alpha$.

In the top right panel of Figure~\ref{fig:bptmodels}, we observe the well-documented offset of $z\sim2.3$ galaxies towards
 higher [O~\textsc{iii}]/H$\beta$ and/or [N~\textsc{ii}]/H$\alpha$ compared to the local star-forming sequence.
  The magnitude of this offset can be observed by comparing the running median of the $z\sim2.3$ sample (dark red line)
 to that of the $z\sim0$ sample (white line).
  However, we do not observe a significant offset between $z\sim0$ and $z\sim2.3$ galaxies in the
 [S~\textsc{ii}] BPT diagram (top left panel), in agreement with results from early MOSDEF data \citep{sha15}.  \citet{mas14} found
 a similar result using a composite spectrum of 24 $z\sim2$ emission-line galaxies from the WISP survey.
  In the [S~\textsc{ii}] BPT diagram, the running median of the $z\sim2.3$ sample is marginally offset to
 lower [O~\textsc{iii}]/H$\beta$ and/or [S~\textsc{ii}]/H$\alpha$, although this offset is not significant given the sample size and
 measurement uncertainty.
  In the middle and lower left panels, it can be seen that an increase in the hardness of the ionizing spectrum
 generally moves grid points to higher [O~\textsc{iii}]/H$\beta$ and [N~\textsc{ii}]/H$\alpha$ at fixed metallicity
 and ionization parameter, which could potentially
 explain the $z\sim2.3$ offset.  In the [S~\textsc{ii}] BPT diagram (middle and lower right), the models show that
 an increase in the hardness of the ionizing spectrum increases [O~\textsc{iii}]/H$\beta$ at fixed [S~\textsc{ii}]/H$\alpha$.
  We emphasize that we are using these models to demonstrate how predicted line ratios change qualitatively as the hardness
 of the ionizing spectrum varies. We are less concerned with a match between a specific blackbody or SB99 spectrum and either
 the local or high-redshift excitation sequences.

In both BPT diagrams, an increase in the hardness of the ionizing spectrum
 leads to a significant increase in [O~\textsc{iii}]/H$\beta$ at fixed [N~\textsc{ii}]/H$\alpha$ or
 [S~\textsc{ii}]/H$\alpha$ for the same ionization parameter and metallicity.
  If a harder ionizing spectrum was the cause of
 the offset of high-redshift galaxies in the [N~\textsc{ii}] BPT diagram, simple photoionization models
 predict that there should also be an offset of similar magnitude towards higher [O~\textsc{iii}]/H$\beta$
 in the [S~\textsc{ii}] BPT diagram.  We observe no significant offset between the $z\sim2.3$ and $z\sim0$ galaxies
 in the [S~\textsc{ii}] BPT diagram.  We note that in the [N~\textsc{ii}] BPT diagram, the 60,000~K
 and 50,000~K blackbody models approximately
 match the position of the $z\sim2.3$ and local galaxies, respectively.
  The same 60,000~K and 50,000~K blackbody models are offset from one another in the [S~\textsc{ii}] BPT diagram,
 and yet no offset is observed in the corresponding observations of $z\sim2.3$ and local galaxies.
  We conclude that there is not a systematically
 harder ionizing spectrum in high-redshift star-forming regions compared to local star-forming regions
 of similar metallicity.  We will revisit the cause of the [N~\textsc{ii}] BPT diagram offset in Section~\ref{sec:bptoffset}.
  Having established that the relation between ionization parameter and O$_{32}$
 is only significantly sensitive to changes in the hardness of the ionizing spectrum, we further conclude
 that the $\mathcal{U}$-O$_{32}$ relation does not strongly evolve with redshift.

\subsubsection{Estimating the typical ionization parameter at $z\sim2.3$}

Having shown that the translation between O$_{32}$ and ionization parameter does not appear to evolve significantly with
 redshift, we will now estimate the range of ionization parameters that would be inferred from the local
 and $z\sim2.3$ galaxy samples.  We advise caution
 when using these ionization parameter estimates because a different ionizing spectrum should be used for the low- and
 high-metallicity galaxy populations and the use of different stellar models will change these estimates.  Matching the
 appropriate stellar models to a given object is further complicated by uncertainty in the metallicity estimates used to tie the
 observed metallicity to the absolute metallicity built into the stellar models.  Since we do not know which input ionizing
 spectrum is appropriate for each sample, we provide estimates of the ionization parameter assuming each of the five ionizing
 spectra considered in the models.  The median $\log($O$_{32})$ value of the $z\sim2.3$
 sample is 0.10 and the middle 68\% span $\log($O$_{32})=-0.11$ to 0.37.
  The local SDSS sample has a median $\log($O$_{32})$ value of $-0.53$ and the middle 68\% span $\log($O$_{32})=-0.71$ to $-0.24$.
  Estimates of the corresponding values of $\log(\mathcal{U})$ are presented in Table~\ref{tab:logu} using the curves in
 Figure~\ref{fig:qo32}.  We note that the median
 ionization parameter is approximately the same for $z\sim0$ and $z\sim2.3$ galaxies if the ionizing spectrum of local
 galaxies is well-described by a 40,000~K blackbody and that of $z\sim2.3$ galaxies is described by a 60,000~K blackbody.
  However, we have shown that $z\sim0$ and $z\sim2.3$ galaxies have similar ionizing spectra at fixed metallicity.  While
 the ionizing spectrum at the median metallicity of each sample will be different, it is unlikely that the magnitude of
 that difference is as large as the difference between a 40,000~K and 60,000~K blackbody spectrum.  Therefore,
 $z\sim2.3$ galaxies have a higher median ionization parameter than local galaxies.

\begin{table*}[t]
 \centering
 \caption{Ionization parameter estimates of local SDSS and $z\sim2.3$ star-forming galaxies based on O$_{32}$
 for each of the five input ionizing spectra assumed in our photoionization models.}\label{tab:logu}
 \begin{tabular*}{0.7\textwidth}{ c @{\extracolsep{\fill}} | l c | c c c c c | }
    &  &  & \multicolumn{5}{c|}{$\log(\mathcal{U})$ assuming an ionizing spectrum of:} \\
    & \multicolumn{2}{c|}{$\log($O$_{32})$} & 40,000 K & 50,000 K & 60,000 K & SB99 soft & SB99 hard \\
  \hline\hline
  \multirow{3}{*}{SDSS} & median & -0.53 & -2.90 & -3.20 & -3.40 & -3.20 & -3.45 \\ \cline{2-8}
    & lower 68\%\tablenotemark{a} & -0.71 & -3.05 & -3.35 & -3.50 & -3.35 & -3.60 \\ \cline{2-8}
    & upper 68\%\tablenotemark{b} & -0.24 & -2.60 & -3.05 & -3.20 & -3.00 & -3.25 \\
  \hline\hline
  \multirow{3}{*}{$z\sim2.3$\ } & median & 0.10 & -2.10 & -2.75 & -2.95 & -2.70 & -3.00 \\ \cline{2-8}
    & lower 68\%\tablenotemark{a} & -0.11 & -2.45 & -2.95 & -3.10 & -2.90 & -3.15 \\ \cline{2-8}
    & upper 68\%\tablenotemark{b} & 0.37 & $\gtrsim-1.5$\tablenotemark{c} & -2.50 & -2.70 & -2.40 & -2.75 \\
  \hline
 \end{tabular*}
 \tablenotetext{1}{The lower bound on the middle 68\% of the distribution of $\log($O$_{32})$}
 \tablenotetext{2}{The upper bound on the middle 68\% of the distribution of $\log($O$_{32})$}
 \tablenotetext{3}{The range of ionization parameters considered in this set of models did not extend high enough to give $\log($O$_{32})=0.37$ for a 40,000 K blackbody input spectrum, so a lower limit is assigned.}
\end{table*}

\subsection{Is the dependence of R$_{23}$ and O3N2 on metallicity redshift invariant?}\label{sec:zo32}

\begin{figure*}
 \centering
 \includegraphics[width=\textwidth]{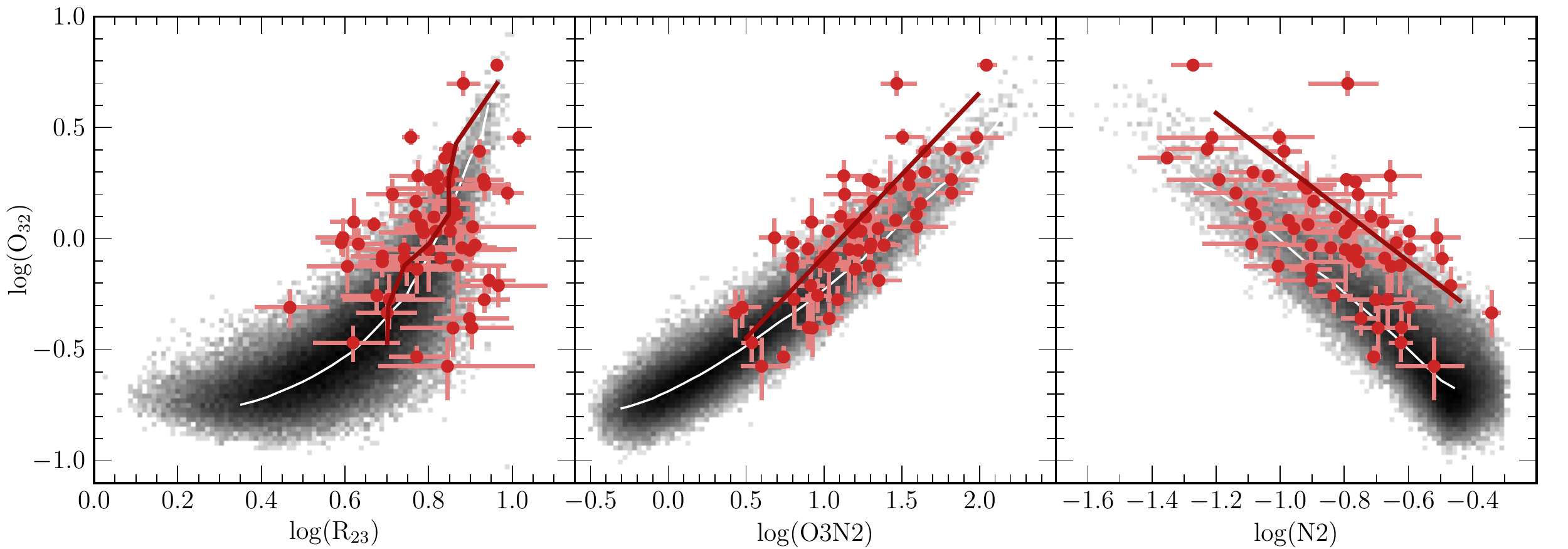}
 \caption{O$_{32}$ vs. R$_{23}$ (left), O3N2 (middle) and N2 (right) for local star-forming galaxies (gray histogram) and
 the $z\sim2.3$ [N~\textsc{ii}] subsample (red points).  In all panels, the white line shows the running median of the local sequence
 of star-forming galaxies.  The dark red line shows the running median of the $z\sim2.3$ sample in the left panel, and the best-fit
 linear relation in the middle and right panels.
}\label{fig:o32combined}
\end{figure*}

We proposed that the relationship between metallicity and ionization parameter
 is the same locally and at $z\sim2.3$.  In Figures~\ref{fig:qo32} and~\ref{fig:bptmodels}, we have shown evidence that the relationship between
 ionization parameter and O$_{32}$ does not significantly change with redshift.  In order for the
 proposed scenario to hold in concordance with Figure~\ref{fig:o32vsr23o3n2}, we must also show that the
 dependence of R$_{23}$ and O3N2 on metallicity does not significantly evolve with redshift.  In
 Figure~\ref{fig:o32combined}, we show the local star-forming sample (gray histogram) and the $z\sim2.3$
 [N~\textsc{ii}] subsample (red circles) in the spaces of
 O$_{32}$ vs. R$_{23}$ (left), O3N2 (middle), and N2 (right).

As previously mentioned, no systematic offset is observed between $z\sim0$ and $z\sim2.3$ galaxies in the
 O$_{32}$ vs. R$_{23}$ diagram, although $z\sim2.3$ objects only occupy the low-metallicity tail of the local
 distribution.  We also pointed out the 0.23 dex offset $z\sim2.3$ galaxies show towards low O3N2
 at fixed O$_{32}$ in Section~\ref{sec:o32metallicity}.  We find a larger systematic offset between
 local and high-redshift galaxies in the O$_{32}$ vs. N2 diagram of 0.33 dex higher N2 at fixed O$_{32}$.
  There is an assymetric scatter towards high N2, such that no $z\sim2.3$ galaxies scatter below
 the local sequence.  We
 additionally observed no significant offset in the [S~\textsc{ii}] BPT diagram (Fig.~\ref{fig:bptmodels}).
  These results collectively suggest that, on average, high-redshift galaxies have higher [N~\textsc{ii}]$\lambda$6584 flux
 compared to the strength of other strong optical emission lines than typical local galaxies with the
 same ionization parameter.  Outside of line ratios involving nitrogen, high-redshift galaxies appear to
 behave similarly to the low-metallicity tail of the local distribution.  We have measurements
 constraining the evolution of density with redshift, and have found no evidence suggesting the
 ionization parameter or hardness of the ionizing spectrum evolve significantly at fixed metallicity.
  Therefore, we conclude that metallicity
 indicators involving nitrogen will evolve with redshift, but the relation between metallicity and location
 in the O$_{32}$ vs. R$_{23}$ diagram is likely the same at $z\sim0$ and $z\sim2.3$.  \citet{jon15} recently
 found that the relationship between R$_{23}$, O$_{32}$, and direct-method metallicity does not evolve
 out to $z\sim0.8$.  \citet{liu08} and \citet{ste14} concluded that the N2 indicator significantly
 over-estimates the metallicity of objects offset from the local star-forming sequence, while the O3N2
 indicator has a much smaller bias.  In
 conjunction with the arguments presented in Section~\ref{sec:qo32}, these results support our proposed scenario
 that galaxies at fixed metallicity have the same ionization parameter locally and at $z\sim2.3$.

\subsection{Additional evidence for the existence of the metallicity-ionization parameter anti-correlation}\label{sec:qohanticorr}

The argument for an anti-correlation between O/H and ionization parameter arises from the fact that
 low-metallicity stars produce more ionizing photons in total and have harder ionizing spectrum \citep{sb99}.
  Observational evidence for this anti-correlation comes from finding that O$_{32}$
 and [O~\textsc{iii}]/H$\beta$ increase with decreasing metallicity \citep{mai08,jon15}.
  However, in Figure~\ref{fig:qo32} it can be seen that harder ionizing spectra give higher values of
 O$_{32}$ at fixed ionization parameter.  We must then consider the possibility that high- and low-metallicity star-forming
 regions have similar ionization parameters while having higher O$_{32}$ values because of the
 change in the hardness of the ionizing spectrum with metallicity.  At fixed $\mathcal{U}$, the range of ionizing spectra
 considered in the models spans $\sim0.9$ dex in O$_{32}$.  Local star-forming galaxies from SDSS span
 a range of $\sim1.9$ dex in O$_{32}$ from $\log($O$_{32})\sim-1.0$ to 0.9 (see Fig.~\ref{fig:o32vsr23o3n2}).
Based on physically-motivated input spectra, it is not possible for the models to span
 a range that large in O$_{32}$ at fixed $\mathcal{U}$.
  It would require spectra that are harder or softer than what is reasonably expected from
 models of the young stellar populations responsible for ionizing HII regions.
  Because the hardness of the ionizing spectrum increases with decreasing metallicity and low-metallicity objects
 are observed to have higher O$_{32}$ values on average, the only way to produce the dynamic range in O$_{32}$
 observed among the SDSS sample is for high-metallicity objects to have low ionization parameters, and vice versa.
  The relation between
 O/H and ionization parameter must exist in the local universe, along with a changing ionizing spectrum with metallicity,
 to reproduce the observed range in O$_{32}$.

Once established, the existence of an anti-correlation between O/H and ionization parameter lends insight
 into the interpretation of the simple models presented in Figure~\ref{fig:bptmodels}.  Using similar
 photionization models to those presented above, \citet{ste14} argued that the sequence of galaxies
 in the [N~\textsc{ii}] BPT diagram at low or high redshift is primarily a sequence in ionization
 parameter because increasing ionization parameter with all other inputs fixed moves grid points along
 the star-forming sequence.  Given the dependence of the ionization parameter and hardness of the
 ionizing spectrum on O/H, many of the grid points shown in Figure~\ref{fig:bptmodels} are not descriptions of real objects.  We do
 not expect to see galaxies with high metallicities, high ionization parameters, and hard ionizing spectra.
  Likewise, galaxies with low metallicities, low ionization parameters, and soft ionizing spectra are not likely
 to be observed.  In fact, it has been previously observed that local HII regions and star-forming galaxies only
 occupy a narrow subset of the parameter space in photoionization model grids \citep[e.g.,][]{dop86,dop06a,dop06b}.
  While individual HII regions can simultaneously demonstrate low metallicity and low ionization parameter,
 as in the sample discussed in \citet{van06}, such objects would contribute negligibly to a luminosity-weighted
 galaxy-averaged spectrum for galaxies that are actively star-forming.

The star-forming sequence in the [N~\textsc{ii}] BPT diagram can be understood as a sequence in both ionization
 parameter and metallicity because the two are fundamentally linked \citep{bre12,sanchez15}.
  \citet{ste14} did mention the relations between ionization parameter, metallicity, and ionizing spectrum
 as a way to reconcile the utility of
 strong-line metallicity indicators in the \textit{local universe} with the apparent lack of dependence of
 the star-forming sequence on metallicity in photoionization models.
  Our results suggest that these relations hold in similar form
 at $z\sim2.3$ as well.  We note that these relations are traced by the \textit{typical} properties of
 the galaxy population, and do not preclude the existence of a small number of galaxies in unexpected regions
 of the parameter space due to scatter in the relations or significantly different conditions caused by some
 rare process or event, such as a major merger.

\subsection{Nitrogen abundance and the cause of the [N~\textsc{ii}] BPT diagram offset}\label{sec:bptoffset}

A significant amount of effort has been put forth to find the cause of the offset
 high-redshift galaxies display in the [N~\textsc{ii}] BPT diagram \citep{kew13t,kew13,mas14,ste14,sha15}.  Proposed
 causes include systematically higher ionization parameters \citep{bri08}, systematically harder ionizing spectra
 \citep{ste14}, elevated nitrogen abundance at fixed metallicity \citep{mas14,sha15}, and higher gas density/ISM pressure \citep{kew13t}.
  It has additionally been proposed that the offset in the BPT diagram is an artifact arising from widespread, weak AGN
 in the high-redshift galaxy population \citep{wri10}.  \citet{coi15} have shown that such global AGN contamination
 does not appear to be present among $z\sim2.3$ galaxies from the MOSDEF survey.

\begin{figure*}
 \centering
 \includegraphics[width=\textwidth]{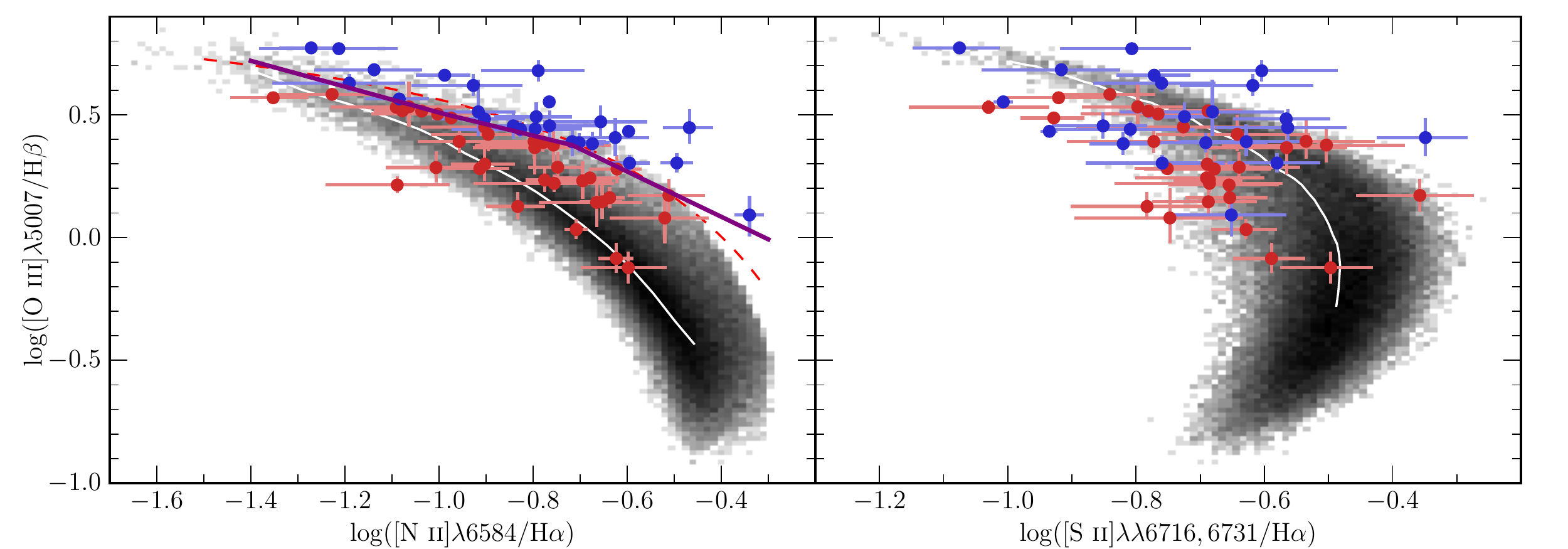}
 \includegraphics[width=\textwidth]{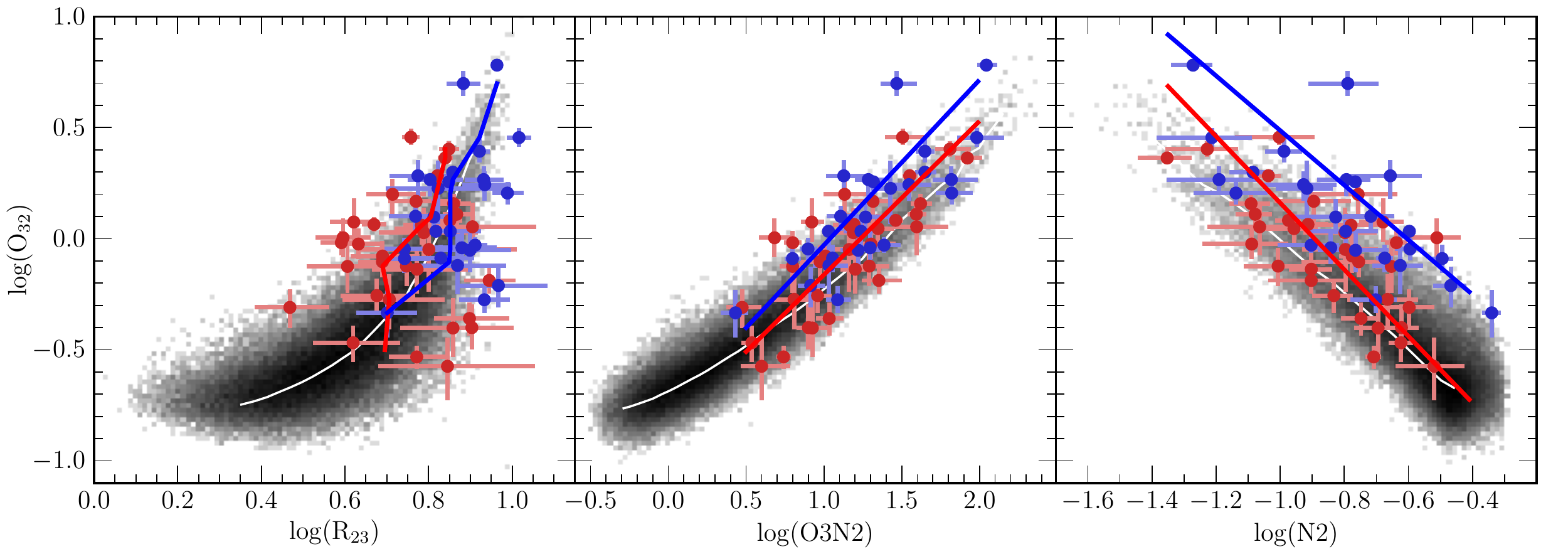}
 \caption{Emission line ratio diagrams with $z\sim2.3$ galaxies color-coded by the offset in the [N~\textsc{ii}]
 BPT diagram.  In each panel, the gray histogram shows the distribution of local star-forming galaxies from SDSS.
  The top row shows the [N~\textsc{ii}] (top left) and [S~\textsc{ii}] (top right) BPT diagrams.
  The O$_{32}$ vs. R$_{23}$, O3N2, and N2 diagrams are shown in the left, middle, and right panels, respectively, of the
 bottom row.
  In each panel, the white line shows the running median of the $z\sim0$ sample.
  In the top left panel, the solid purple line shows the running median of the $z\sim2.3$
 star-forming galaxies, while the dashed red
 line shows the fit to the MOSDEF $z\sim2.3$ star-forming sequence from \citet{sha15}.
  In all panels, $z\sim2.3$ galaxies falling above and to the right of the dashed red line are color-coded blue, while $z\sim2.3$
 galaxies falling below and to the left of this line are color-coded red.
  In the lower left panel, the red and blue curves show the median R$_{23}$ in bins of O$_{32}$ for the red and blue data points, respectively.
  The solid red and blue lines in the lower middle and right panels
 show the best linear fit to the red and blue data points, respectively.
  See Footnote 9 regarding the use of running medians and best-fit lines.
}\label{fig:colorcoded}
\end{figure*}

In this paper, we have sought to characterize the physical properties influencing the ionization state
 of high-redshift galaxies, and can now use the results presented herein to investigate the cause of
 the [N~\textsc{ii}] BPT diagram offset.  Figure~\ref{fig:colorcoded} shows the [N~\textsc{ii}] and [S~\textsc{ii}]
 BPT diagrams (Fig.~\ref{fig:bptmodels}) and the O$_{32}$ vs. R$_{23}$, O3N2, and N2 diagrams
 (Fig.~\ref{fig:o32combined}) with the $z\sim2.3$ galaxies color-coded according to the magnitude of the offset
 in the [N~\textsc{ii}] BPT diagram.  We divided the $z\sim2.3$ sample in the [N~\textsc{ii}] BPT diagram at the
 running median [O~\textsc{iii}]/H$\beta$ in bins of [N~\textsc{ii}]/H$\alpha$,
 shown by the solid purple line in the top right panel.
  Galaxies above and to the right of this line in the [N~\textsc{ii}] BPT diagram  are shown in blue,
 while those galaxies that show a smaller offset and overlap the
 local sequence are shown in red.  We have verified that our results do not change if other
 dividing lines in the [N~\textsc{ii}] BPT diagram are used, including a linear fit and the best-fit line
 to the $z\sim2.3$ star-forming sequence from \citet{sha15} (red dashed line).

We find that the blue and red data points are well mixed and are not systematically offset from the local
 distribution in the O$_{32}$ vs. R$_{23}$ diagram.
  The running median of the three samples follow a very similar sequence where there is overlap
 at $\log($O$_{32})\sim-0.1$ to 0.4.  Similarly,
 the blue data points do not appear to follow a
 different distribution from that of the red data points or the low-metallicity local sequence in the [S~\textsc{ii}]
 BPT diagram.  There is some separation in the two subsamples in the [S~\textsc{ii}] BPT diagram because the blue data points have
 higher [O~\textsc{iii}]/H$\beta$ a priori due to the selection.
  In the O3N2 diagram, the red points fall close to the local distribution, while the blue points appear to be slightly more offset
 towards lower O3N2 at fixed O$_{32}$.  The blue data points display less overlap with the red data points in the N2 diagram and
 a larger offset compared to local galaxies.
  We use the best-fit linear relations to the blue and red datapoints, respectively, to quantify the mean offsets and make the
 offsets clearer.  The red line has a mean offset of -0.11 dex in O3N2 at fixed O$_{32}$ compared to the local median, while the
 blue line is offset -0.30 dex on average.  In the N2 diagram, the red and blue lines are offset 0.08 and 0.40 dex, respectively,
 towards higher N2 at fixed O$_{32}$.

To summarize, we find that, as a whole, the $z\sim2.3$ sample shows no significant offset in line ratio diagrams that
 do not include nitrogen ([S~\textsc{ii}] BPT and O$_{32}$ vs. R$_{23}$ diagrams).  We find a slight offset
 in a diagram that utilizes a line ratio including nitrogen as well as oxygen (O$_{32}$ vs. O3N2 diagram).  This
 offset increases in the N2 diagram, in which nitrogen is the only metal species in the line ratio.  Additionally,
 we find that galaxies showing the largest offset in the [N~\textsc{ii}] BPT diagram do not distinguish themselves
 from galaxies showing a smaller offset when plotted in diagrams that do not include nitrogen.  The most-offset galaxies in the
 [N~\textsc{ii}] BPT diagram display
 systematically lower O3N2 values at fixed O$_{32}$ than galaxies showing a small offset in the [N~\textsc{ii}] BPT diagram.  This
 effect increases in magnitude in the O$_{32}$ vs. N2 diagram, in which those galaxies showing a large offset in the
 [N~\textsc{ii}] BPT diagram display significantly larger N2 values at fixed O$_{32}$ than the less-offset subsample
 of the $z\sim2.3$ galaxies.  A simple way to change the [N~\textsc{ii}]$\lambda$6584 flux without affecting the
 flux of the other strong optical emission lines is to change the nitrogen abundance.
  Our observations suggest that the N/O ratio at fixed O/H is higher on average in $z\sim2.3$ star-forming galaxies
 compared to local star-forming galaxies.  The high-redshift galaxies showing the largest offset in the
[N~\textsc{ii}] BPT diagram appear to have higher N/O ratios than less offset $z\sim2.3$ galaxies of the same
 metallicity.

\begin{figure}
 \centering
 \includegraphics[width=\columnwidth]{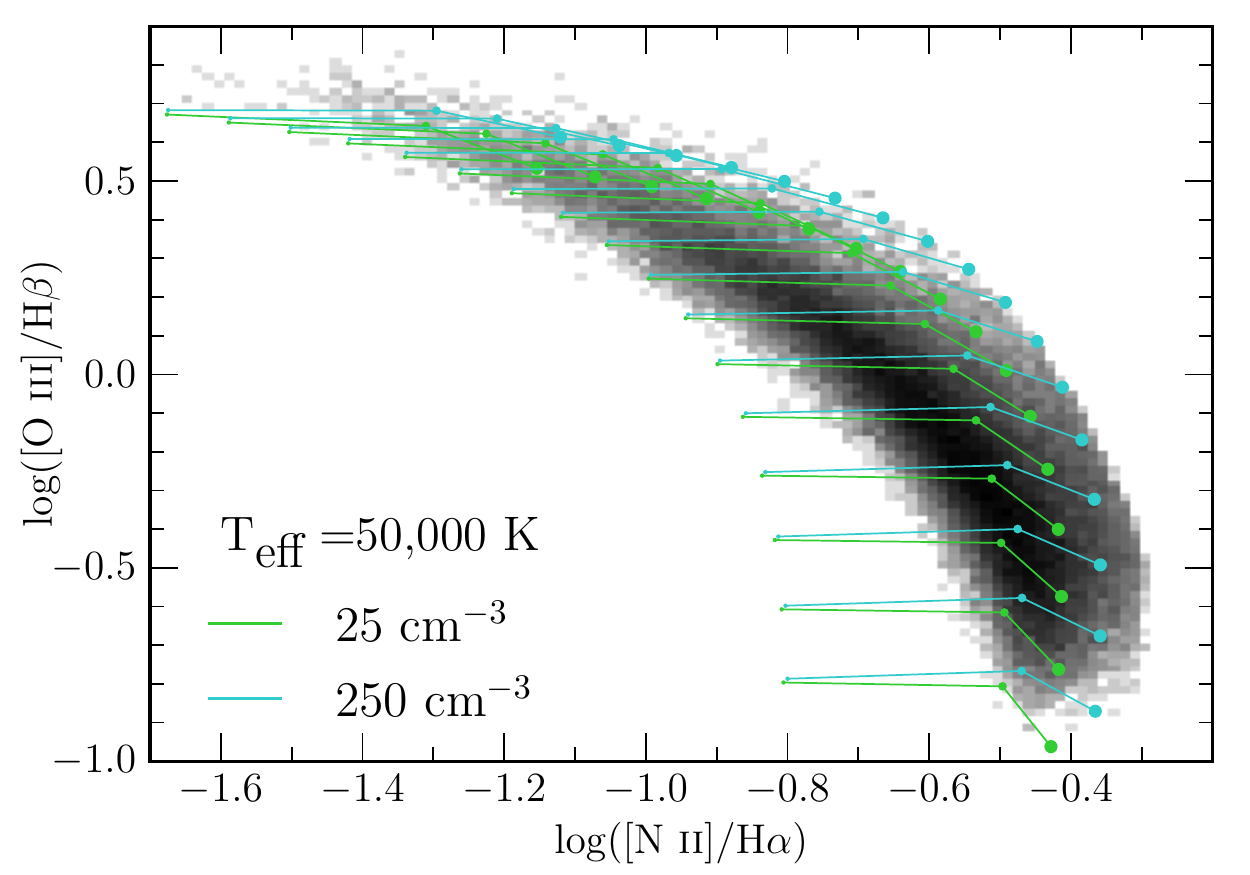}
 \caption{Model grids assuming a gas density of 25~cm$^{-3}$ (light green) and 250~cm$^{-3}$ (cyan) with
 gas-phase metallicity spanning 0.2-0.6~Z$_{\odot}$ and an input ionizing spectrum of a 50,000~K blackbody.
  Grid points are displayed as in Figure~\ref{fig:bptmodels}.
}\label{fig:bptdens}
\end{figure}

In Sections~\ref{sec:qo32} and \ref{sec:zo32}, we showed that our data are inconsistent with a
 systematically harder ionizing spectrum or higher ionization parameter in high-redshift galaxies at
 fixed metallicity.  An increase in the gas density or ISM pressure can also move galaxies in the direction
 of the observed offset.  We utilize Cloudy photoionization models to quantify the magnitude of this effect
 using the characteristic densities obtained in Section~\ref{sec:dens}.  Figure~\ref{fig:bptdens} shows the
 local star-forming sequence for reference, along with two photoionization model grids produced with the same
 ionizing spectrum (blackbody with T$_{\rm eff}=50,000$ K) while varying the electron density between 25 cm$^{-3}$
 and 250 cm$^{-3}$.  The grids are shown over the metallicity range 0.2-0.6 Z$_\odot$ in which the typical metallicity
 at $z\sim2.3$ is expected to fall.  The lowest metallicity grid points show negligible change in line ratios with
 density, while the dependence on density increases with metallicity.  With an increase of a factor of 10 in density,
 [O~\textsc{iii}]/H$\beta$ and [N~\textsc{ii}]/H$\alpha$  are increased by $\lesssim0.1$ dex at fixed ionization
 parameter and metallicity.  This is not
 a large enough shift to account for the observed offset in the [N~\textsc{ii}] BPT diagram.
  If our density estimate for local star-forming regions is underestimated,
 then the magnitude of the line ratio shift caused by density will be even smaller.
  Thus, the increase in density 
 from $z\sim0$ to $z\sim2.3$ plays only a minor role in the [N~\textsc{ii}] BPT diagram offset.

We conclude that the observed offset of $z\sim2.3$ galaxies from the local star-forming sequence in the
 [N~\textsc{ii}] BPT diagram is mostly caused by elevated N/O at fixed O/H compared to local
 galaxies, while an increase in density/pressure of the star-forming regions plays a minor role.  We do
 not find evidence that a change in the ionizing spectrum or ionization parameter at fixed metallicity
 plays a part in the offset.  This finding is consistent with earlier MOSDEF results from \citet{sha15} attributing
 the offset in the BPT diagram to higher N/O in high-redshift galaxies with \mstar$<10^{10}$~M$_{\odot}$.
  Our results are also in agreement with the interpretation of \citet{mas14}, in which anamolous
 nitrogen abundance was first proposed as a cause of the offset.  \citet{ste14} also found an
 elevated N/O among their $z\sim2.3$ sample, but argued that the primary cause of the offset lies in
 harder ionizing spectra and higher ionization parameters in comparison to what is observed locally.

The cause of the higher N/O ratios observed at high-redshift is not yet clear.  \citet{mas14}
 speculated that a larger-than-normal population of Wolf-Rayet stars could produce a nitrogen
 enhancement, although a mechanism to produce such a population exclusively at high redshifts was not proposed.
  If the nitrogen enhancement is due to a significant difference in the stellar population at a given
 metallicity, there would likely be a difference in the overall ionizing spectrum that population produces,
 for which we do not see evidence.  Another possibility is that gas flows lead to a larger nitrogen abundance
 at fixed O/H.  If a large amount of unenriched gas is accreted and mixed into the star-forming regions
 before intermediate-mass stars enrich the ISM with nitrogen, O/H will decrease.  The intermediate-mass stars
 will then release an amount of nitrogen that is larger than what is expected at that O/H because the gas
 is now less enriched than the stars.  This scenario, or another mechanism involving gas inflows and outflows,
 is perhaps more likely in high-redshift galaxies which are known to have large gas fractions and concentrated star
 formation that could correspond to active accretion of metal-poor gas \citep{tac13}.
  We found evidence consistent with an accumulation of unenriched gas in $z\sim2.3$ galaxies
 through the mass-metallicity relation in \citet{san15}.  Observations of the rest-frame UV could determine
 if exotic stellar populations are present, while more observations of the cold gas content of high-redshift
 galaxies with, e.g., ALMA could help uncover the role of gas flows.

The presence of AGN within the $z\sim2.3$ sample could significantly bias conclusions drawn from excitation diagrams.
  We did not use a locally-determined demarcation to separate star-forming galaxies from AGN in the [N~\textsc{ii}]
 BPT diagram \citep[e.g.,][]{kew01,kau03b} because such a selection would introduce a bias against objects with high
 N/O at fixed [O~\textsc{iii}]/H$\beta$.  While some of the galaxies in the $z\sim2.3$ sample fall in the local AGN region
 of the [N~\textsc{ii}] BPT diagram, we do not see any $z\sim2.3$ galaxies present where the AGN sequence would fall
 in the other four panels of Figure~\ref{fig:colorcoded}.  Therefore, we are confident that our $z\sim2.3$ sample is free
 of AGN.

\subsection{Diffuse ionized gas and the interpretation of global galaxy spectra}\label{sec:diffuse}

The observed positions of local and $z\sim2.3$ galaxies in the [N~\textsc{ii}] and [S~\textsc{ii}] BPT diagrams are key
 pieces of evidence in our arguments regarding the hardness of the ionizing spectrum of local and high-redshift star-forming regions
 (see Section~\ref{sec:qo32} and Figure~\ref{fig:bptmodels}).  One source of uncertainty in the interpretation of these plots
 is the inclusion of light from the diffuse ionized component of the ISM in integrated-light galaxy spectra.
  Emission from diffuse ionized gas can be significant, with roughly half of the total H$\alpha$ emission of local spiral
 galaxies coming from a diffuse component \citep{zur00}.  If the diffuse ionized component has line ratios that are not
 equivalent to those of HII regions, which is probable because diffuse gas and HII regions are characterized by
 different ionizing sources and
 ionization states, then light from the diffuse component can act as a contaminant when investigating properties of
 star-forming regions.  This issue is especially concerning for low ionization states such as [N~\textsc{ii}] and [S~\textsc{ii}]
 which are easily ionized and can have elevated flux compared to that of H$\alpha$ when shock-excited \citep{mar97,hon13}.  Characterizing
 the impact of diffuse emission requires knowledge of both the fraction of emission and the emission line ratios coming
 from the diffuse ionized component, at both low and high redshift.

In the local universe, comparison of integrated galaxy spectra with those of individual HII regions can give insight into
 the effect of emission from other components of the ISM.  Local HII regions have been found to follow a similar sequence
 to that of local star-forming galaxies in the [N~\textsc{ii}] BPT diagram, but appear to have systematically
 lower [S~\textsc{ii}]/H$\alpha$ at fixed [O~\textsc{iii}]/H$\beta$ in the [S~\textsc{ii}] BPT diagram \citep{vei87,pil12,ber15,cro15}.
  Every emission line
 in an integrated galaxy spectrum contains a luminosity-weighted contribution from each HII and diffuse emission region falling
 in the slit or fiber, complicating the comparison of HII region and integrated galaxy spectra.  Ongoing spatially-resolved
 spectroscopic surveys, such as the SDSS-IV/MaNGA IFU \citep{manga} and SAMI \citep{bry15} galaxy surveys, will provide a dataset capable of unraveling
 the relative contributions of HII regions and diffuse ionized gas to global galaxy spectra.

The situation is much more uncertain at high redshifts because the structure of the ISM in $z\sim2$ galaxies is poorly constrained.
  The ISM is almost certainly different in $z\sim2$
 galaxies,  considering the high gas fractions \citep{tac13} and compact sizes \citep{van14}, combined with high rates of
 star formation \citep{shi15}, significant feedback driving outflows \citep{sha03,ste10}, and high level of turbulence
 observed in $z\sim2$ disk galaxies \citep{for09}.
  If high-redshift galaxies are dominated by giant kpc-scale HII regions
 ionized by super-star clusters that fill a large fraction of the galaxy volume, the filling factor of the diffuse
 ionized component may be small along with the fraction of line emission originating there.
  In this case, it may be more appropriate to compare the integrated emission-line spectra of high-redshift galaxies
 to those of luminous local HII regions rather than global galaxy spectra.
  We are only starting to gain rudimentary knowledge of the spatially-resolved structure of the ionized ISM at $z\sim2$
 \citep{for11,gen11,jon13,new14}.
Adaptive optics observations of increased sensitivity and spatial resolution will be required to map the spatially-resolved
 structure of the ionized component of the ISM at high redshift, including the strength of the [N~\textsc{ii}] and
 [S~\textsc{ii}] emission lines.

\section{Summary}\label{sec:summary}

We have investigated the physical conditions of star-forming regions at $z\sim2.3$, specifically the
 electron density and ionization parameter, and made comparisons to local star-forming galaxies in order
 to understand how these properties evolve with redshift.  We performed this investigation using rest-frame
 optical spectra of $z\sim2.3$ galaxies from the ongoing MOSDEF survey.  We summarize our main conclusions below and
 discuss future observations that could shed additional light on the ionization state of high-redshift galaxies.

\begin{enumerate}
\item We explored the evolution of the electron density in star-forming regions using the
 [O~\textsc{ii}]$\lambda\lambda$3726,3729 and [S~\textsc{ii}]$\lambda\lambda$6716,6731 doublets.
  We found that $z\sim2.3$ galaxies have median [O~\textsc{ii}]$\lambda3729/\lambda3726=1.18$ and
 median [S~\textsc{ii}]$\lambda6716/\lambda6718=1.13$, corresponding to electron densities of 225~cm$^{-3}$
 and 290~cm$^{-3}$, respectively.
  Local star-forming galaxies from SDSS have median
 [S~\textsc{ii}]$\lambda6716/\lambda6718=1.41$ which yields a density of  $26$~cm$^{-3}$.
  We found an evolution in electron density of an order of magnitude between $z\sim0$ and $z\sim2.3$.
\item We investigated the ionization state of $z\sim2.3$ and local star-forming galaxies, using O$_{32}$ as a proxy
 for the ionization parameter.  We found that O$_{32}$ decreases with increasing stellar mass in both the local and $z\sim2.3$ samples.
  The slope of the O$_{32}-$\mstar\ anti-correlation is very similar for both samples, but the $z\sim2.3$ sample
 is offset $\sim0.6$ dex higher in O$_{32}$ relative to the local sample at fixed \mstar.  This offset can be explained
 by the evolution of the mass-metallicity relation with redshift, such that high-redshift galaxies have lower metallicities
 at fixed \mstar, and the existence of an anti-correlation between ionization parameter and O/H.
\item We found that $z\sim2.3$ galaxies show no systematic offset from local galaxies and follow the distribution of
 the low-metallicity tail of local galaxies in the O$_{32}$ vs. R$_{23}$ diagram.  The high-redshift sample behaves similarly
 to the local sample in the O$_{32}$ vs. O3N2 diagram, displaying a slight systematic offset from the local
 distribution.  We propose that $z\sim2.3$
 galaxies follow the same anti-correlation between ionization parameter and O/H that is observed in the local universe.
\item Using simple photoionization models, we demonstrated that the translation between O$_{32}$ and ionization parameter
 is only strongly dependent on the shape of the ionizing spectrum, and has little dependence on the assumed gas density and
 gas-phase metallicity. This translation will only evolve with redshift if the hardness of the ionizing spectrum at fixed
 metallicity evolves with redshift.
\item We utilized the position of $z\sim2.3$ and local galaxies in the [N~\textsc{ii}] and [S~\textsc{ii}] BPT diagrams
 combined with simple photoionization models to show that the hardness of the ionizing spectrum does not significantly
 increase or decrease with redshift.  Photoionization models predict that a hardening of the ionizing
 spectrum will increase [O~\textsc{iii}]/H$\beta$ at fixed [N~\textsc{ii}]/H$\alpha$ and [S~\textsc{ii}]/H$\alpha$.
  The $z\sim2.3$ sample displays an offset from the local sequence in the [N~\textsc{ii}] BPT diagram but is not significantly offset
 in the [S~\textsc{ii}] BPT diagram.  We conclude that there is not a significant increase in the hardness of the
 ionizing spectrum at fixed metallicity between $z\sim0$ and $z\sim2.3$.
\item Galaxies at $z\sim2.3$ show no significant systematic offset from local galaxies in line ratio diagrams involving only lines of oxygen,
 sulfur, and hydrogen, while they show a systematic offset in line ratio diagrams involving nitrogen.  These results suggest that
 metallicity indicators using line ratios excluding nitrogen (e.g. the combination of O$_{32}$ and R$_{23}$) do not
 evolve up to $z\sim2.3$, while indicators using nitrogen are biased due to an evolution in N/O at fixed O/H.
\item A consequence of conclusions 6 and 7 is that $z\sim2.3$ have similar ionization parameters to $z\sim0$ galaxies at
 fixed metallicity.  Higher typical ionization parameters are inferred for $z\sim2.3$ galaxies compared to those of local
 galaxies because $z\sim2.3$ galaxies have lower typical metallicities.  The ionization state appears to be set by the metallicity
 both locally and at $z\sim2.3$.
\item We investigated the offset between $z\sim2.3$ and local galaxies in the [N~\textsc{ii}] BPT diagram.  We found that the
 $z\sim2.3$ galaxies that display the largest offsets in the [N~\textsc{ii}] BPT diagram are not significantly offset
 from the local distribution or the remainder of the $z\sim2.3$ sample in the O$_{32}$ vs. R$_{23}$ and [S~\textsc{ii}]
 BPT diagrams, but are systematically offset in the O$_{32}$ vs. O3N2 and N2 diagrams.  We conclude that higher N/O at fixed
 O/H drives the $z\sim2.3$ offset in the [N~\textsc{ii}] BPT diagram.  We previously provided evidence against significant evolution
 of the hardness of the ionizing spectrum or ionization parameter at fixed O/H.  We used simple photoionization models
 to show that an evolution of a factor of 10 in the gas density cannot account for the full offset in the [N~\textsc{ii}] BPT
 diagram.  We further conclude that
 an increase in the gas density plays a minor secondary role in driving the $z\sim2.3$ offset in the [N~\textsc{ii}] BPT diagram.
\end{enumerate}

There still remain many questions to be answered regarding the ionization state of local and $z\sim2.3$ galaxies.  Emission
 line contribution from the diffuse ionized component of the ISM is uncertain both locally and at high redshifts.  Variety in photoionization
 modeling leads to variation in estimated ionization parameters.  There are some observations that would
 allow us to test our proposed scenario that $z\sim2.3$ galaxies have the same ionization parameter as $z\sim0$ galaxies with the same
 metallicity.
  The most obvious of these is direct-method oxygen abundance measurements from auroral lines at $z\sim2$.
  Direct-method abundances are insensitive to changes in the electron density, showing variations of $\lesssim0.01$ dex from 25~cm$^{-3}$
 to 250~cm$^{-3}$ (i.e., the observed evolution in density from $z\sim0$ to $z\sim2.3$), an advantage over some strong-line methods
 \citep[see][]{jon15}.
  Currently, only a handful of direct-method abundance measurements have been attained at $z\gtrsim1$, often utilizing gravitational lensing
 \citep{vil04,yua09,erb10,rig11,bra12b,chr12,mase14}.  Current sensitive near-infrared spectrographs on 10~m class telescopes and
 upcoming instruments on the Thirty Meter Telescope will allow for observations of auroral lines for more typical galaxies
 at $z>1$.  Additionally, observations of other ionization-parameter-sensitive emission line ratios could provide a test of
 this scenario, including [Ne~\textsc{iii}]$\lambda$3869/[O~\textsc{ii}]$\lambda\lambda$3726,3729 \citep{lev14} and
 [S~\textsc{iii}]$\lambda\lambda$9069,9532/[S~\textsc{ii}]$\lambda\lambda$6716,6731 \citep{kew02}.  Finally, improved stellar and
 photoionization models that can make self-consistent predictions of the physical properties of star-forming galaxies across the entire
 local star-forming sequence are needed to create more reliable ionization parameter calibrations.
  Such models would provide a local foundation for reliable photoionization models of the full suite of strong rest-frame
 optical emission lines out to $z\sim2$.

\acknowledgements We thank Tucker Jones for enlightening conversations, and acknowledge the First Carnegie Symposium
 in Honor of Leonard Searle for useful information and discussions that benefited this work.
 We acknowledge support from NSF AAG grants AST-1312780, 1312547, 1312764, and 1313171, and archival grant AR-13907,
 provided by NASA through the Space Telescope Science Institute.
 We are also grateful to Marc Kassis at the Keck Observatory for his many valuable contributions to
the execution of the MOSDEF survey. We also acknowledge
the 3D-HST collaboration, who provided us with spectroscopic
and photometric catalogs used to select MOSDEF targets and derive
stellar population parameters. We also thank I. McLean, K. Kulas,
and G. Mace for taking observations for the MOSDEF survey in May and June 2013.
MK acknowledges support from the Hellmann Fellows fund. ALC
acknowledges funding from NSF CAREER grant AST-1055081. NAR is supported by
an Alfred P. Sloan Research Fellowship.
We wish to extend special thanks to those of Hawaiian ancestry on
whose sacred mountain we are privileged to be guests. Without their generous hospitality, most
of the observations presented herein would not have been possible.
 This work is also based on observations made with the NASA/ESA Hubble Space Telescope, which is operated by the
 Association of Universities for Research in Astronomy, Inc., under NASA contract NAS 5-26555. Observations associated
 with the following GO and GTO programs were used: 12063, 12440, 12442, 12443, 12444, 12445, 12060, 12061, 12062, 12064
 (PI: Faber); 12177 and 12328 (PI: van Dokkum); 12461 and 12099 (PI: Riess); 11600 (PI: Weiner); 9425 and 9583
 (PI: Giavalisco); 12190 (PI: Koekemoer); 11359 and 11360 (PI: O’Connell); 11563 (PI: Illingworth).

\bibliography{densityionpaper}

\end{document}